\documentclass[11pt,floatfix,showpacs]{revtex4}
\usepackage[dvips]{graphicx,color}
\usepackage{amsmath}
\usepackage{amsfonts}
\usepackage[latin1]{inputenc}

\begin{document}

\title{Quark matter under strong magnetic fields in SU(2) NJL-type models:
parameter dependence of the cold dense matter phase diagram}

\author{Pablo G. Allen$^{a}$ and N.N.\ Scoccola$^{a,b,c}$}

\address{$^{a}$ Physics Department, Comisi\'on Nacional de
Energ\'{\i}a Atómica, Av.Libertador 8250, 1429 Buenos Aires,
Argentina. \\
$^{b}$ CONICET, Rivadavia 1917, 1033 Buenos Aires, Argentina. \\
$^{c}$ Universidad Favaloro, Sol{\'\i}s 453, 1078 Buenos Aires,
Argentina.}

\begin{abstract}
The phase structure of magnetized cold quark matter is analyzed in
the framework of the two-flavor Nambu-Jona-Lasinio models paying
special attention to its dependence on the model parameters as
different values within the phenomenological allowed range are
considered. We first discuss the simpler chiral limit case, and
then the more realistic situation of finite current masses. We
show that in spite of the difference in the nature of some
transitions, both cases are alike and exhibit a rather rich phase
structure for a significant range of acceptable parameters. A
simplification of the phase structure is obtained as parameters
leading to larger values of the dressed quark mass in the vacuum
are considered. Finally, we consider the so-called ``inverse
catalysis effect" showing that in some phases it implies an actual
decrease of the order parameter as the magnetic field increases.
\end{abstract}


\maketitle

\section{Introduction}

Understanding the behavior of strongly interacting matter under
the influence of intense magnetic fields has become an issue of
increasing interest in recent years \cite{Kharzeev:2013jha}. This
has been mostly motivated by the realization that strong magnetic
fields may be produced in several physically relevant situations.
For example, present estimates \cite{Kharzeev:2007jp} indicate
that in non-central heavy ion collisions at very high energies the
field intensity could be $B \sim 10^{19}$ G, i.e. $eB \sim 0.06\
\mbox{GeV}^2$ in natural units. Moreover, the compact stellar
objects believed to be the source of intense $\gamma$ and X rays,
magnetars, are expected to bear fields of the order of
$10^{13}-10^{15}$ G at their surface reaching values several
orders of magnitude greater at their center \cite{Duncan:1992hi}.
Note that in all these situations the matter is, in addition,
subject to extreme conditions of temperature and/or density. Thus,
it is of great interest to investigate which modifications are
induced by the presence of strong magnetic fields on the whole QCD
phase diagram. Unfortunately, even in the absence of those fields,
the present knowledge of such phase diagram is only schematic.
Only recently have powerful lattice QCD (LQCD) simulations
\cite{Borsanyi:2010bp} firmly established that for 2+1 flavors and
vanishing baryon chemical potential there is a crossover-like
transition at $T_{pc} \simeq 160$~MeV from a hadronic phase, in
which chiral symmetry is broken and quarks are confined, to a
partonic phase in which chiral symmetry is restored and/or quarks
are deconfined. The situation is less clear for finite chemical
potentials due to the well-known difficulty given by the so-called
sign problem which affects lattice calculations
\cite{Karsch:2003jg}. Of course, the presence of strong magnetic
fields makes the situation even more complex. Thus, most of our
present knowledge of their effect comes from investigations
performed in the framework of effective models (see e.g.
Refs.\cite{Fraga:2012rr,Gatto:2012sp} and refs. therein). A
general outcome is an enhancement at vanishing chemical potential
of the dynamical symmetry breaking due to external magnetic field,
a phenomenon usually referred to as ``magnetic
catalysis"\cite{Gusynin:1994re}. In fact, a recent  LQCD
study\cite{Bali:2011qj} of the behavior of the $u-$ and $d-$
condensates at zero and finite temperature in an external magnetic
field has confirmed the magnetic catalysis phenomena predicted by
most of the models at zero temperature. However, for temperatures
of the order of the crossover temperature a decrease of the quark
condensates is found. It remains an open and interesting question
what prevents magnetic catalysis to persist for these larger
temperatures. In this article we will concentrate on a different
sector of the phase diagram: that of finite chemical potential and
low temperatures. Although this region has been the subject of
several investigations in the past (see e.g. Refs.
\cite{Ebert:1999ht,Ebert:2003yk,Inagaki:2003yi,Menezes:2008qt,Boomsma:2009yk,Fayazbakhsh:2010gc,Ferrari:2012yw,Ferrer:2012wa}),
as in the case of vanishing magnetic field the corresponding
behavior of the strongly interacting matter has not been firmly
established. For example, only very recently was it fully realized
that there exists an ``inverse catalysis effect" at certain values
of the magnetic field \cite{Preis:2010cq}. We will perform our
analysis in the framework of the two-flavor Nambu-Jona-Lasinio
(NJL)-type models \cite{reports}. As is well known (see
e.g.Ref.\cite{Buballa:2003qv}), even in the simplest version of
these models there is a significant range of phenomenologically
acceptable values for the model parameters. In this situation,
previous NJL studies of the effect of the magnetic field on cold
quark matter have only considered some particular choices of
allowed parameterizations. Our aim is to perform a systematic
analysis of how both the qualitative and quantitative details of
the phase diagram of cold dense quark matter subject to intense
magnetic fields depend on the specific choice of the parameters.
It should be noted that the generic features of such phase diagram
have been first studied in Refs.\cite{Ebert:1999ht,Ebert:2003yk}.
However, in these works only the chiral limit was considered and
no details of the precise dependence of the phase diagram on
parameterizations within the range of phenomenological interest
were given. As already mentioned, for the more realistic case of
finite current quark masses only a few particular
parameterizations were considered.

This work is organized as follows. In Sec.\ II we provide a brief
review of the NJL model description of cold dense quark matter in
the presence of external magnetic fields. The model parameters, as
well as the way to determine them, are also introduced. In Sec.\
III we consider the parameter dependence of the phase diagrams in
the chiral limit. The case of finite quark masses is analyzed in
Sec.\ IV. Our main conclusions are presented in Sec.\ V. Finally,
we have included two Appendices: in App. A we provide some details
of the parameterizations for the chiral case while in App. B we
give the numeric values of parameters used for the case of finite
current masses.

\section{Formalism}

Our starting point is the Euclidean effective action of the SU(2) NJL model in the presence
of an external electromagnetic field. It reads:
\begin{equation}
S_E = \int d^4 x
\left\{
\bar \psi (- i \gamma^{\mu} D_\mu + m_c) \psi
- G\left[(\bar \psi \psi)^2  +  (\bar \psi i\tau \gamma_{5}\psi)^2 \right]
\right\},
\end{equation}
where $m_c$ is the current quark mass (we work in the isospin
limit $m_c=m_u=m_d$)  and $G$ is a coupling constant. The coupling
of the quarks to the electromagnetic field ${\cal A}_\mu$ is
implemented through the covariant derivative $D_{\mu}=\partial_\mu
- i q_f {\cal A}_{\mu}$ where $q_f$ represents the quark electric
charge ($q_u/2 = -q_d = e/3$). We consider a  static and constant
magnetic field in the $z$ direction, ${\cal A}_\mu=\delta_{\mu 2}
x_1 B$. Since the model under consideration is not renormalizable,
we need to specify a regularization scheme. Here, we introduce a
sharp cutoff in 3-momentum space, only for the ultra-violet
divergent integrals. Together with $m_c$ and $G$, the cutoff
$\Lambda$ forms a set of three parameters that completely
determine the model. These parameters are usually fixed so as to
reproduce the empirical values in the vacuum of the pion mass
$m_\pi$, the pion decay constant $f_\pi$, and the quark condensate
$<\bar q q >_0$. The latter is related to dressed quark mass in
the vacuum $M_0$ via $M_0= m_c - 2 G <\bar q q>_0$. Whereas the
physical values $m_\pi = 138.0$ MeV and $f_\pi = 92.4$ MeV are
known quite accurately, the uncertainties for the quark condensate
are rather large. Limits extracted from sum rules are $190
\mbox{MeV} < -<u \bar u>_0^{1/3} < 260$ MeV at a renormalization
scale of 1 GeV\cite{Dosch:1997wb}, while typical lattice
calculations yield $-<u \bar u>_0^{1/3} =  231 \pm 8 \pm 6$ MeV
\cite{Giusti:1998wy} (see e.g. Ref.\cite{McNeile:2005pd} for some
other lattice results). As a consequence of this, different
parametrizations compatible with this rather broad range of values
for the condensate have been used in the literature. As frequently
done, here we choose to take $M_0$ as the quantity which defines
those parametrizations.  To be compatible with the above mentioned
phenomenological values for the quark condensate we must have $300
\lesssim M_0 \lesssim 600$ MeV \cite{Buballa:2003qv}.

To account for finite temperature $T$ and chemical potential $\mu$
one can follow the standard Matsubara formalism which amounts to
performing the replacements,
\begin{eqnarray}
p_4 \rightarrow (2 n+ 1) \pi T - i \mu
\qquad ; \qquad
\int \frac{dp_4}{2 \pi} \rightarrow \sum_n
\end{eqnarray}
In the case of the local NJL model under consideration the sum
over Matsubara modes can be analytically performed. Since we are
particularly interested in analyzing the behavior of cold quark
matter, we take the limit of vanishing temperature in the
resulting expressions. In this limit, the thermodynamical
potential in the Mean Field Approximation (MFA)
reads\cite{Ebert:2003yk,Menezes:2008qt}
\begin{eqnarray}
\Omega(\mu,B, M)&=& \frac{(M-m_c)^2}{4G} + \frac{N_cN_f}{8
\pi^2}\left\{ M^4 \ln \frac{\Lambda + \sqrt{\Lambda^2+M^2}}{M} -
\Lambda (2 \Lambda^2 + M^2) \sqrt{ \Lambda^2 + M^2}  \right\}
\nonumber\\
&-& \!\!\! \frac{N_c}{2\pi^2} \! \sum_{f=u,d}(q_f B)^2
\left[ \zeta'(-1,x_f) + \frac{x_f^2}{4} - \frac{1}{2} (x_f^2 -x_f)\ln x_f \right]
\nonumber\\
&-& \!\!\!  \frac{N_c}{4\pi^2}  \! \sum_{k,f}  \  \theta(\mu - s_{kf}) \ \alpha_k |q_f| B
\left\{ \mu \sqrt{\mu^2 - s_{kf}^2} -
s_{kf}^2 \ln \left[ \frac{ \mu + \sqrt{\mu^2 - s_{kf}^2 }}{s_{kf}} \right]
\right\}
\label{omega}
\end{eqnarray}
where $\alpha_k = 2- \delta_{k0}$, $s_{kf}=\sqrt{M^2 + 2 k |q_f|
B}$ and $x_f= M^2/(2 |q_f| B)$. In addition, $\zeta'(-1,x_f) = d
\zeta(z,x_f)/dz |_{z=-1}$ where $\zeta(z,x_f)$ is the
Riemann-Hurwitz zeta function. The sum over $k$ in the last line
corresponds to the sum over the populated Landau levels (LL's)
associated to each quark flavor $f$. The dressed quark mass
$M(\mu,B)$ at a given value of $\mu$ and $B$ is found as solution
of the gap equation, $\partial\Omega/\partial M = 0$. As is
well-known, the behavior of these solutions as a function of $\mu$
indicates the existence of some kind of phase transition which, in
the chiral case, can be of first or second order depending on the
value of the external magnetic field. For finite current masses,
however, these possible second order phase transitions become
smooth crossovers. Consequently, there is not a unique way to
define their position. In fact, even in the absence of a magnetic
field, different prescriptions have been used in the literature to
define the position of a crossover-type transition. They include,
for example, the location of the peak of the chiral susceptibility
$\chi_{ch}= \partial <\bar q q >/\partial m_{c}$, of the peak of
the derivative of an order parameter with respect to some
thermodynamical variable (as e.g. $\mu$ or $T$), etc. This issue
will be discussed in some detail in Sec.IV. We should also mention
that, as will be seen in Sec.III, some additional second order
transitions can occur in the chirally restored phase $M=0$ in the
chiral case. There, one should consider some additional quantity,
like the quark number density $\rho = -\partial \Omega/\partial
\mu$, in order to observe their effect.

As already stated, the main aim of this work is to perform a
detailed analysis of how the character and location of these
different types of phase transitions depend on the chosen
parametrization of the NJL model. For convenience, we discuss in
the next section the simplified case of $m_c=0$. The more
realistic case of finite quark mass will be addressed in Sec.IV.

\section{Phase diagrams in the chiral limit}

In this section we analyze the chiral case $m_c=0$. As discussed
in the Appendix~A, in this case the model has only two parameters:
the coupling constant $G$ and the 3D cutoff $\Lambda$. In order to
work along a line of ``constant physics" we determine them as a
function of $M_0$ so as to reproduce a value of the pion decay
constant in the chiral limit $f_\pi^{ch}=90$ MeV. The numerical
results for the dimensionless coupling constant $g=G\Lambda^2$ and
$\Lambda$ as functions of $M_0$ are given in the Appendix~A (see
upper panel of Fig.\ref{figapp}). Note that since for $m_c=0$ the
pion decay constant is the only dimensionful quantity in the
problem, any dimensionful quantity (expressed in natural units)
has to be the product of some $f_\pi^{ch}$-independent constant
multiplied by some power of it. In this way, all the results to be
shown in this section can be easily made ``universal", in the
sense of being independent of the chosen value for $f_\pi^{ch}$.
Of course, some extra dependence on the chosen procedure to
regularize the UV divergencies might still exist.

\begin{figure}[t]
\hspace*{3cm}
\includegraphics[width=0.5\textwidth]{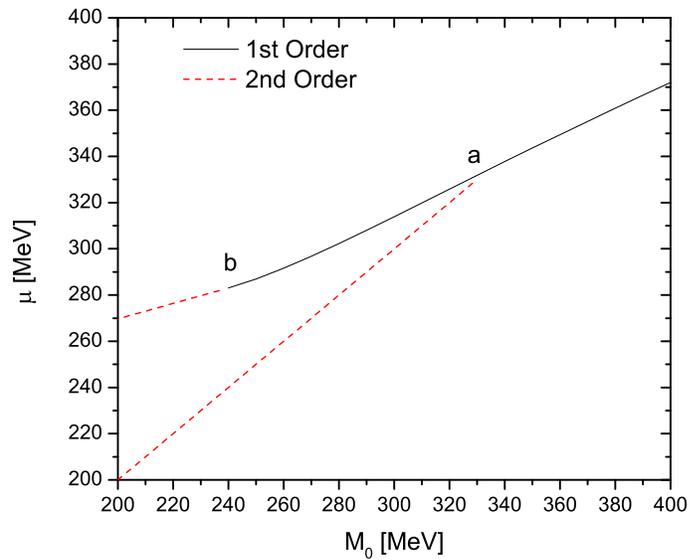}
\hspace*{\fill} \caption{(Color online) Critical chemical
potentials as functions of the model parametrization (specified by
the value of $M_0$)
    in the absence of the magnetic field.} \label{fig1}
\end{figure}

\begin{figure}[b]
\includegraphics[width=0.5\linewidth,angle=0]{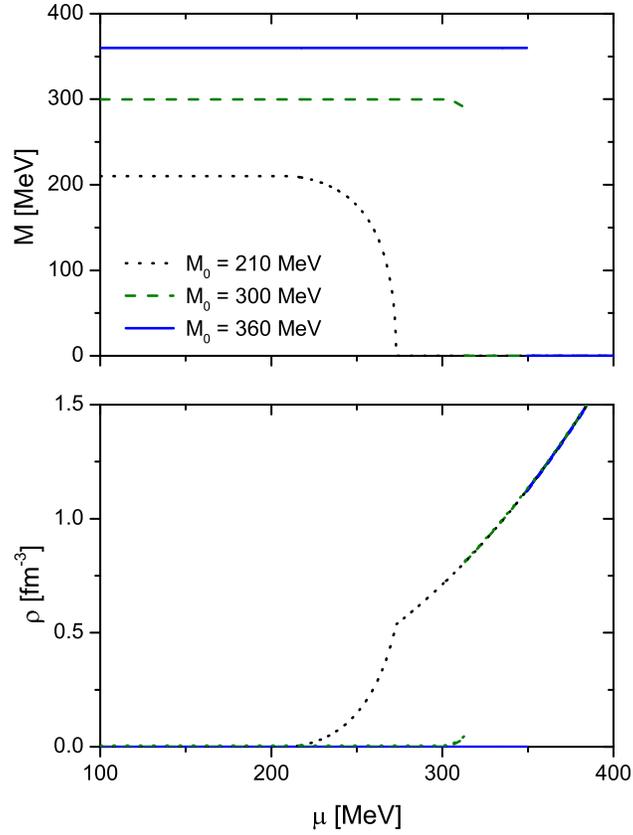}
\caption{(Color online) Behavior of the dressed mass $M$ (upper
panel) and the quark density $\rho$
    (lower panel) for $eB=0$. Plots for several representative model
    parameter sets specified by the value of $M_0$ are shown.}
\label{fig2}
\end{figure}

We start by discussing the situation at vanishing magnetic field.
Although this has already been discussed in the literature to some
extent \cite{Buballa:1996tm,Klimenko:1997nn,Vshivtsev:1997pj} it
will serve as a benchmark for a better understanding of the
modifications introduced by the presence of external magnetic
fields. The corresponding diagram in the $M_0-\mu$ plane is shown
in Fig.\ref{fig1}. There, the full lines correspond to first order
phase transitions while the dashed lines to second order ones. Let
us recall that each value of $M_0$ corresponds to a different
parametrization of the model. As we see, depending on the value of
$M_0$, three different regions can be distinguished. For $M_0 <
M_0(b)$ two consecutive second order transitions occur as $\mu$
increases. The first one connects a phase in which $M=M_0$
independently of the value $\mu$ to one in which $M=M(\mu,0)$ with
$0 < M < M_0$. The second transition in turn connects the latter
phase to the chirally restored phase $M=0$. Note that while the
first of these transitions implies a discontinuity of $d^2 M/d
\mu^2$  at the critical point \cite{Klimenko:1997nn}, in the case
of the second one already the first derivative is discontinuous
 at the corresponding point. For $M_0(b) < M_0 <
M_0(a)$ the situation is similar except for the fact that the
second transition is of first order type. Finally, for $M_0 >
M_0(a)$ there is only one first order transition connecting the
phase with $M=M_0$ to the one with $M=0$. These possible
situations are illustrated in Fig.\ref{fig2} where we display the
behavior of $M$ (upper panel) and the density $\rho$ (lower panel)
as a function of $\mu$ for three values of $M_0$, each one lying
in one of the above mentioned regions. For our chosen value of
$f_\pi^{ch}$ we obtain $M_0(a) = 334.45$ MeV and $M_0(b) = 239.64$
MeV. As already indicated, these quantities can be written in a
``universal" way if we express them in terms of $f_\pi^{ch}$. We
obtain
\begin{equation}
M_0(a) = 3.716 f_\pi^{ch} \qquad ; \qquad M_0(b) = 2.663
f_\pi^{ch}
\end{equation}
It should be mentioned that an approximation to the above
expression for $M_0(a)$ was given in Ref.\cite{Buballa:1996tm}
where the relation $M_0(a) \simeq 4 f_\pi^{ch}$ was quoted. On the
other hand, in Ref.\cite{Ebert:2003yk} the values of $M_0(a)$ and
$M_0(b)$ were given in terms of the cutoff $\Lambda$. Although in
principle correct, we find that this way to express these critical
masses might be somewhat inconvenient.  Note that for a fixed
ratio $M_0/\Lambda$, different values of $\Lambda$ correspond to
different values of $f_\pi^{ch}$ (see Eq.(\ref{fpich})) and, thus,
do not represent the same physical situation. For this reason  we
find it more adequate to take $f_\pi^{ch}$ instead of $\Lambda$ as
the ``independent" variable.

\begin{figure}[t]
    \includegraphics[width=0.9\linewidth,angle=0]{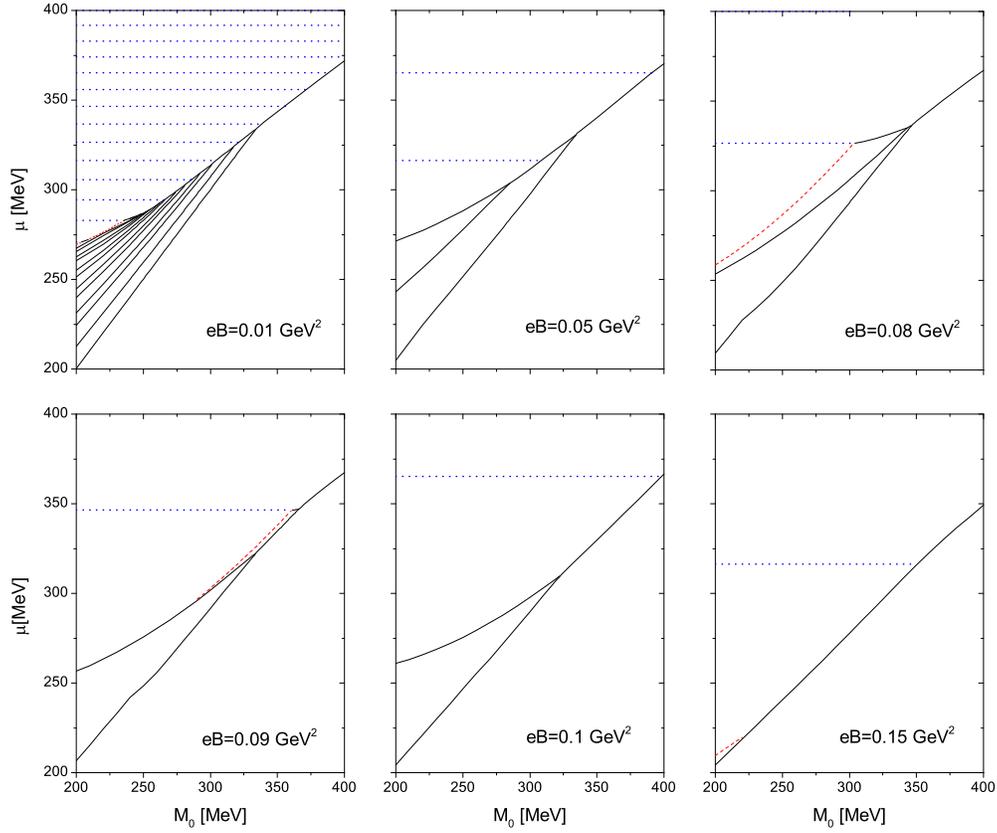}
    \caption{(Color online) Critical chemical potentials as functions of the model parametrization specified by the value of $M_0$
    for several representative values of the magnetic field.}
\label{fig3}
\end{figure}

\begin{figure}[t]
    \includegraphics[width=0.6\linewidth,angle=0]{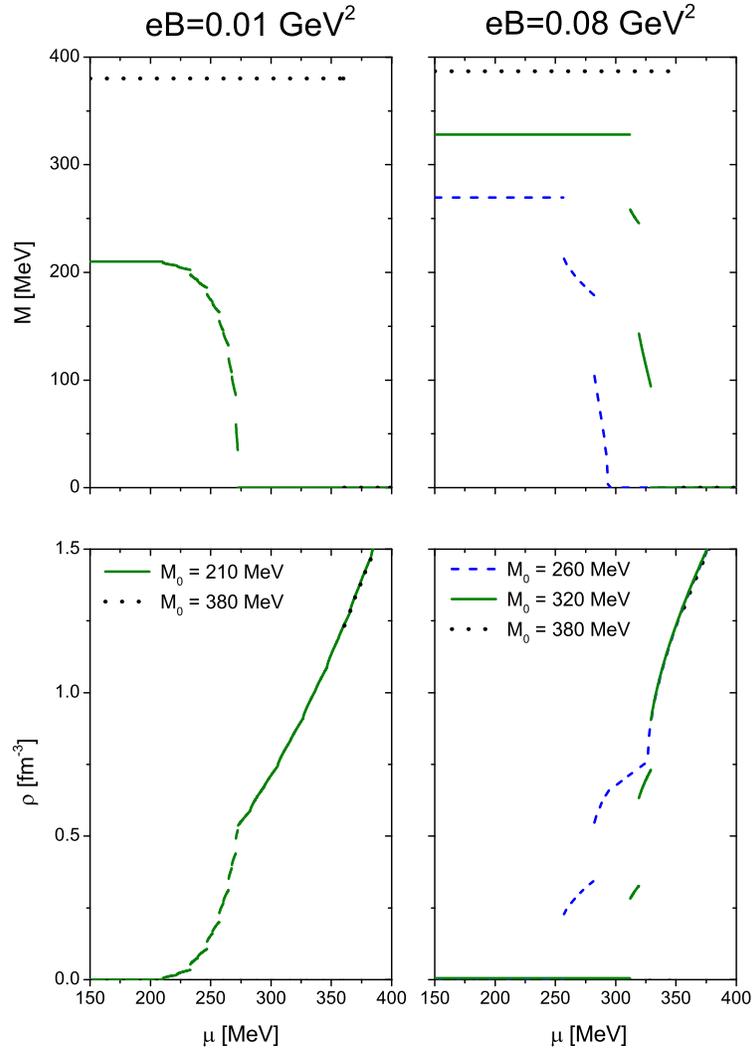}
    \caption{(Color online) Behavior of the dressed mass $M$ (upper panels) and the quark density $\rho$
    (lower panels) for two selected values of $eB$ and several representative model parameter
    sets specified by the value of $M_0$.}
\label{fig4}
\end{figure}

We turn now to the finite $B$ case. In Fig.\ref{fig3} we display
the behavior of the critical chemical potentials of the different
possible transitions as functions of $M_0$ for several
representative values of $eB$. Note again that if the values of
$\mu$, $M_0$ and $eB$ are scaled with the corresponding powers of
$f_\pi^{ch}= 90$ MeV these figures are ``universal" in the sense
described above. We concentrate first in the lowest value of the
magnetic field considered, $eB=0.01 \ \mbox{GeV}^2$. Comparing it
with the corresponding one for $eB=0$ shown in Fig.\ref{fig1}, we
observe some differences and similarities. Firstly, here we can
observe that there is a clearly different behavior depending on
whether $M_0 > M_0(a)$ or not. Note that in general the value of
$M_0(a)$ depends on $eB$ (see Fig.5 below and its corresponding
discussion). For parametrizations with $M_0 > M_0(a)$ there is
only one first order transition at a given critical $\mu$. Such
transition connects the chirally broken phase with $M=M(0,B)$ to
the chiral phase $M=0$. On the other hand for $M_0 < M_0(a)$ the
chirally restored phase is reached only after a succession of
several first order transitions. This situation is illustrated in
the left panels of Fig.\ref{fig4} where we display the behavior of
$M$ (upper panel) and the quark density $\rho$ (lower panel) for
$eB=0.01 \ \mbox{GeV}^2$ and two representative values of $M_0$.
It is interesting to analyze the case $M_0 < M_0(a)$ in some
detail. Contrary to what happens for $eB=0$, where the lowest
critical $\mu$ corresponds to a second order transition, we note
that as soon as a small external field is present such transition
becomes first order. It is possible to check that if one takes,
for example, $M_0 = 210$ MeV and considers values of $eB < 0.01 \
\mbox{GeV}^2$, the number of first order transitions needed to
reach the $M=0$ phase increases as $eB$ decreases and, eventually,
the curves of $M$ and $\rho$ as functions of $\mu$ tend to those
shown in Fig.\ref{fig2} as $eB$ vanishes. It is clear that the
discontinuities present in the finite $eB$ case are due to the
quantization of the (Landau) levels induced by the magnetic
fields. Another important observation concerns the last transition
before the $M=0$ phase is reached. Such transition can be of first
or second order depending on the chosen value of $M_0$. Although
somewhat difficult to observe in the case of $eB = 0.01 \
\mbox{GeV}^2$, this effect becomes clear as larger values of $eB$
are considered. Finally we note that, independently of the chosen
parametrization, in the chirally restored phase $M=0$ extra second
order transitions occur at the chemical potentials indicated by
the dotted lines.  The corresponding critical values of $\mu$ are
such that some new Landau levels contribute to the last line of
Eq.(\ref{omega}) for $M=0$. Thus,
\begin{equation}
\mu = \frac{2}{3}\ k \ c_f \ eB
\label{va}
\end{equation}
where $c_u = 2 c_d = 2$. It is important to note that, although in
the chiral case these transitions do not cause any change in the
chiral order parameter, some other quantities like the derivatives
of the quark density do display discontinuities at the critical
point. This effect can be observed, for example, in the behavior
of the density as function of $\mu$ for the case $M_0=210$ MeV
displayed in the left lower panel of Fig.\ref{fig4}. In fact, all
the features present there are associated with the magnetic
oscillations related to the so-called ``van Alphen - de Haas
effect"\cite{Ebert:1999ht}. Note that while for chemical
potentials leading to massive phases these oscillations induce
first order transitions, for those corresponding to massless ones
the transitions are of second order.

Continuing with the analysis of the $M_0 - \mu$ diagrams displayed
in Fig.\ref{fig3} we discuss now how they are modified as $eB$
increases. We see that for $eB = 0.05 \ \mbox{GeV}^2$ the value of
$M_0(a)$ is somewhat larger than for $eB = 0.01 \ \mbox{GeV}^2$.
In addition, for a given $M_0 < M_0(a)$ the number of first order
transitions needed to go from the $M=M(0,B)$ phase to the one with
$M=0$ decreases. Of course, as is clear from Eq.(\ref{va}), the
spacing between the dotted lines becomes larger. Note also that,
for the range of values of $M_0$ considered, all the transitions
connecting the finite $M$ phases to the vanishing $M$ ones are of
first order type. The situation changes for  $eB = 0.08 \
\mbox{GeV}^2$ since in this case there is a region of values of
$M_0$ for which the $M=0$ phase is reached through a second order
phase transition. It is interesting to note that such second order
line ends at the point where a dotted line meets a first order
line. As seen in previous cases, however, the intersection of a
dotted line and a first order one does not necessarily imply the
existence of a second order ``chiral" line that ends at the
meeting point. The behavior of $M$ and $\rho$ as functions of
$\mu$ for some representative values of $M_0$ are shown in the
right panels of Fig.\ref{fig4}. For the next value considered, $eB
= 0.09 \ \mbox{GeV}^2$, we observe that the point $a$ still moves
towards larger values of $M_0$, and both the beginning and the end
of the second order line are inside the considered range of
parametrizations. However, if we increase the magnetic field
further to $eB = 0.10 \ \mbox{GeV}^2$ the position of $a$ displays
a sudden jump towards smaller values of $M_0$. Moreover, no second
order ``chiral" line is present for $M_0~>~200$ MeV. In the last
case explicitly considered, which corresponds to $eB = 0.15 \
\mbox{GeV}^2$, we see that the point $a$ has moved further to the
lower left corner of the diagram. In addition a new second order
line appears, but only for parametrizations corresponding to small
values $M_0 < 220$ MeV. We note that for even larger values of
$eB$ we get $M_0(a) < 200$ MeV and, thus, there is only one first
order transition in the whole range of parametrizations
considered. Of course, in addition to such transition, there also
exist the corresponding second order transitions which are always
present in the chirally restored phase.

\begin{figure}[t]
\hspace*{3cm}
    \includegraphics[width=0.6\linewidth,angle=0]{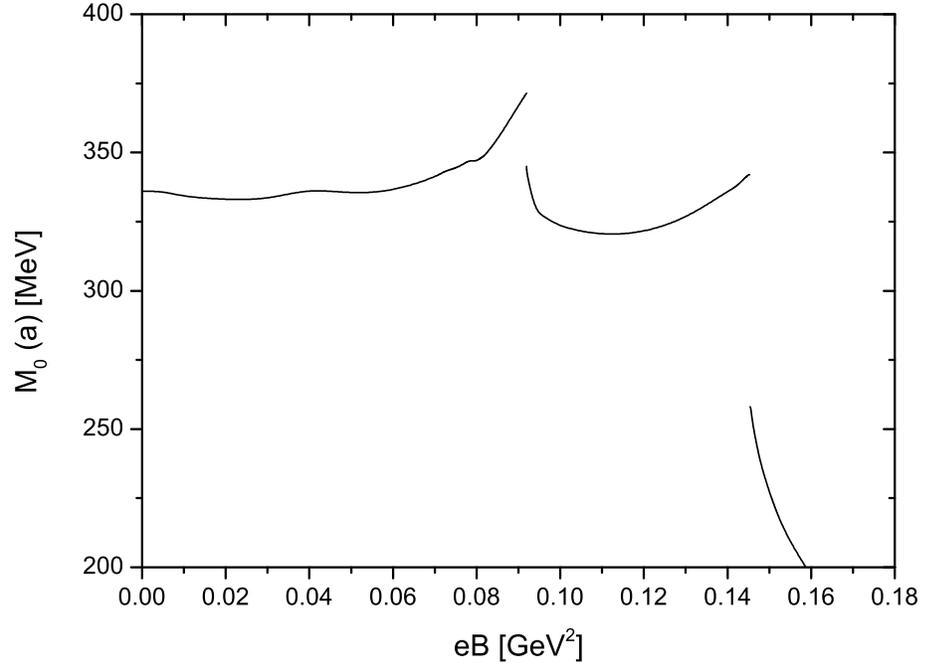}
    \hspace*{\fill}
    \caption{Position of the critical point $a$ as a function of $eB$ in the chiral case.
    Note that for parametrizations with $M_0 > M_0(a)$ only one first order transition connects
    the vacuum phase to the chirally restored phases.}
\label{fig5}
\end{figure}

From the analysis above it is clear that the presence of the
magnetic field induces a rather rich and diverse structure of
phase transitions for the different possible parametrizations
laying within the physical range  $300 \lesssim M_0 \lesssim 600$
MeV. Particularly interesting is the behavior of the position of
the critical point $a$ as a function of $eB$. In fact, for
parametrizations with $M_0 > M_0(a)$ only one first order phase
transition connects the vacuum phase to the chirally restored
phase. Such behavior is shown in Fig.\ref{fig5}. As we see, the
corresponding curve is not monotonic and presents two peaks
followed by two associated discontinuities. Expressed in
``universal" fashion the highest maximum ($hm$) corresponds to
\begin{equation}
M_0(a)^{hm} = 4.127 \ f_\pi^{ch}\qquad  \mbox{with}\qquad eB^{hm}
= 11.35 \ (f_\pi^{ch})^2 \label{mhm}
\end{equation}
while the lowest maximum ($lm$) corresponds
\begin{equation}
M_0(a)^{lm} = 3.798 \ f_\pi^{ch}\qquad \mbox{with}\qquad eB^{lm} =
17.95 \ (f_\pi^{ch})^2
\end{equation}
which for our chosen value $f_\pi^{ch} = 90$ MeV leads to
$M_0(a)^{hm}= 371.46$ MeV and $M_0(b)^{lm}=341.80$ MeV. We should
mention that, in principle, additional peaks and discontinuities
might appear for larger values of $eB$. However the corresponding
values of $M_0(a)$ would be below $200$~MeV and, thus, far from
the physical range of interest. The value of $M_0(a)^{hm}$ is
particularly important. In fact, for parameterizations with $M_0 >
M_0(a)^{hm}$ the phase diagram in the $eB-\mu$ plane is very
simple since, as it will be shown below, it only displays a single
first order phase transition for all values of the magnetic field.
It is interesting to note that the interval $M_0(a_{eB=0}) < M_0 <
M_0(a)^{hm}$ is the one mentioned in footnote 9 of
Ref.\cite{Ebert:2003yk} for which stable quark droplets are formed
by massive quarks. Note, however, that no precise value for
$M_0(a)^{hm}$ was given in that reference.

\begin{figure}[t]
\hspace*{2cm}
    \includegraphics[width=0.6\linewidth,angle=0]{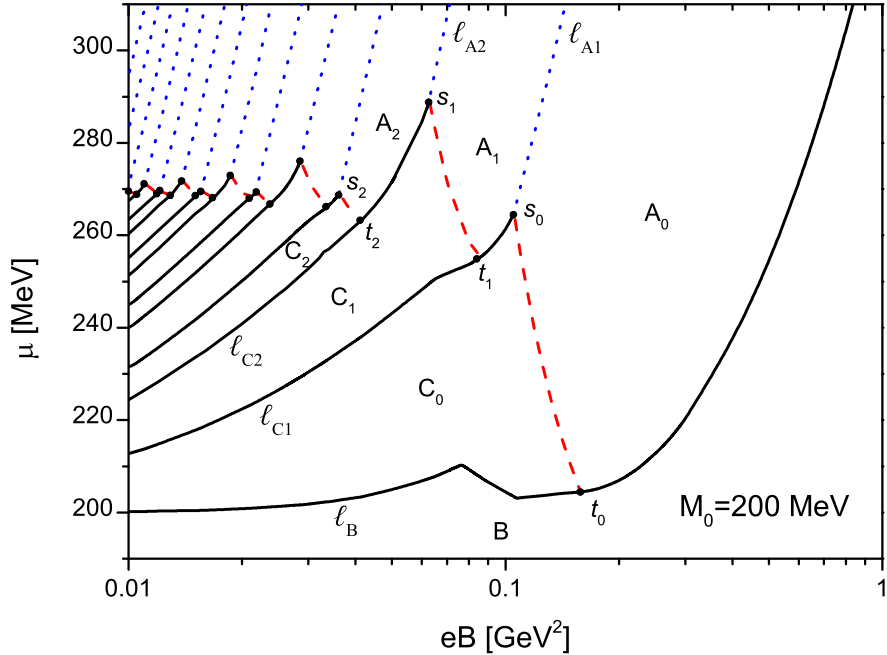}
\hspace*{\fill}
    \caption{(Color online) Phase diagrams in the $eB-\mu$ plane for the chiral case and
    $M_0 = 200$ MeV. Full (black) lines represent first order phase transitions
    while dashed (red) lines represent second order ones. Dotted (blue) lines correspond to the second
    order transitions which separate the different massless phases.
    Phase B corresponds to the fully chiral symmetry broken phase with no LL populated, while
    the phases $\mbox{C}_i$ to massive phases in which LL's up to $k=i$ for
    $d$-quarks
    and $k=m$ for $u$-quarks are populated.
    Here, $m=i/2\ (\ (i-1)/2\ )$ if $m$ is even (odd).
    In phase B, the dressed mass takes the vacuum value $M(0,eB)$ independently of $\mu$
    while in
    the phases $\mbox{C}_i$ it takes a smaller value which does depend on $\mu$. The phases
    $\mbox{A}_i$ are phases in which chiral symmetry is restored and LL's up to $k$
    (related to $i$ as above) are populated.}
\label{fig6}
\end{figure}

\begin{figure}[t]
    \includegraphics[width=0.9\linewidth,angle=0]{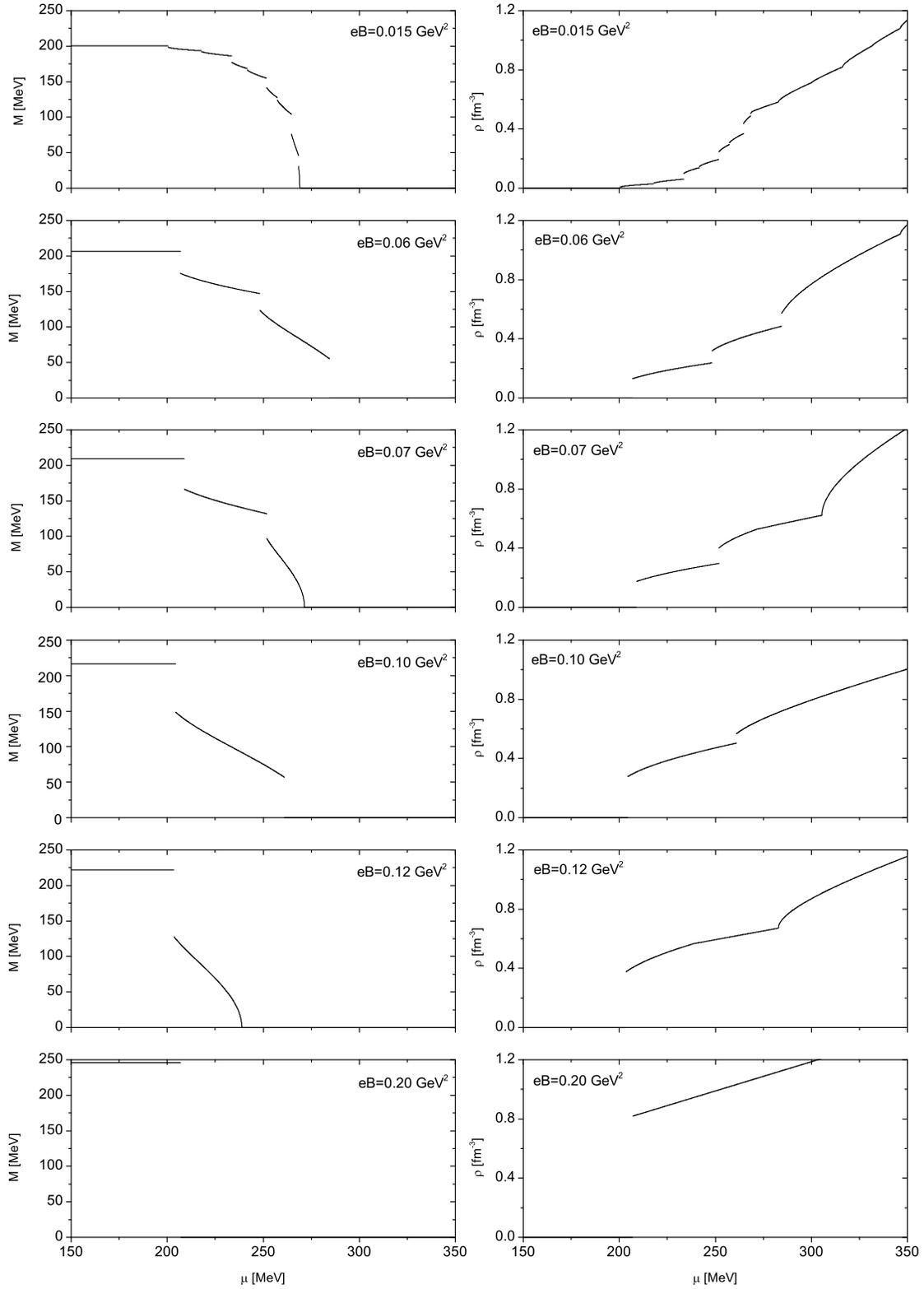}
    \caption{Behavior of the dressed mass $M$ (left panels) and the quark density (right panels) as
    functions of the chemical potential for the chiral case with $M_0 = 200$ MeV and several representative
    values of the magnetic field.}
\label{fig7}
\end{figure}

We turn now to the analysis of the $eB-\mu$ phase diagrams.
However, before considering values of $M_0$ within the accepted
range of physical interest, we will focus on the situation for
$M_0= 200$~MeV. Although this might be only considered a case of
academic interest, it is nevertheless instructive since it
displays the full complexity that a phase diagram of this type
might have, allowing also to appreciate how such diagram is
simplified as $M_0$ increases towards the physical region of
parameters. The corresponding phase diagram is shown in
Fig.\ref{fig6}. This diagram is very similar to the one sketched
in Fig.4 of Ref.\cite{Ebert:1999ht} which corresponds to a
simplified one-flavor model. Note that we use different types of
lines to represent the various types of transitions. In
Fig.\ref{fig6}, full (black) lines correspond to first order phase
transitions, dashed (red) lines to second order ``chiral"
transitions and dotted (blue) lines to the second order
transitions between massless phases. On the other hand we follow
the notation of Ref.\cite{Ebert:1999ht} to denote the different
phases and critical points. In fact, the phase B corresponds to
the fully chirally broken phase where $M=M(0,B)$, while the phases
$\mbox{C}_i$ correspond to massive phases where $M$ also depends
on $\mu$. This can be clearly observed in Fig.\ref{fig7} where we
display the behavior of $M$ (left panels) and $\rho$ (right
panels) for different representative values of $eB$. Note that
while the quark density in the $B$ phases vanishes, this is not
the case in the phases $\mbox{C}_i$. Finally, the phases
$\mbox{A}_i$ correspond to massless phases with different number
of populated LL's. For convenience, we also introduce the
following notation for the lines separating the different phases.
We use $\ell_{\text{B}}$ to indicate the first order line that
separates the B phase from the $\mbox{C}_0$ or $\mbox{A}_0$
phases, ${\ell}_{\text{C}i}$ (with $i=1,2..$) the first order line
separating the $\mbox{C}_i$ and $\mbox{C}_{i-1}$ phases and
${\ell}_{\text{A}i}$ (with $i=1,2..$) the second order line
separating the $\mbox{A}_i$ and $\mbox{A}_{i-1}$. Note, however,
that in general there exists a segment of the ${\ell}_{\text{C}i}$
line (that going from $t_i$ to $s_{i-1}$) that actually separates
the $\mbox{A}_i$ and $\mbox{C}_{i-1}$ phases. As discussed in
Ref.\cite{Ebert:1999ht}, for vanishing $eB$ all the
${\ell}_{\text{C}i}$ lines are expected to meet
${\ell}_{\text{B}}$ at a single point, M, with
$\mu(\mbox{M})=M=200$ MeV in the present case. Note that this
point should lie on the lower dashed line of Fig.\ref{fig1} and,
thus, corresponds to a second order transition point. The main
difference with Fig.4 of Ref.\cite{Ebert:1999ht} is the form of
the segments connecting the points $t_i$ and $s_{i-1}$. In fact,
we find that the slope of the corresponding functions $\mu(eB)$ is
always positive and increases with $eB$. The equations for the
${\ell}_{\text{A}i}$ lines are given by Eq.(\ref{va}). Note that
each time that one of these lines is crossed from right to left
some new LL's are populated. In fact, the crossing of
${\ell}_{\text{A}1}$ corresponds to the population of the
$d$-quark state with $k=1$, that of ${\ell}_{\text{A}2}$ to the
{\it simultaneous} population of the $u$-quark state with $k=1$
and $d$-quark state with $k=2$, etc. The fact that the population
of the $u$-quark state with a certain $k$ coincides with the one
of the $d$-quark state with $2 k$ is simply due to the fact that
(in modulus) the electric charge of the first is twice that of the
second. Thus, those ${\ell}_{\text{A}i}$ associated with odd $i$
correspond to population of only one $d$-quark state while the
ones with even $i$ to the simultaneous population of a $d$-quark
and a $u$-quark. Consequently larger effects are expected to
happen when crossing the ``even" ${\ell}_{\text{A}i}$. A similar
phenomenon occurs when crossing the first order
${\ell}_{\text{C}i}$ lines. This pattern can be particularly well
observed in the upper panels of Fig.\ref{fig7}. Here one should
keep in mind that the first transition (the one with the lowest
$\mu$) corresponds to the crossing of the ${\ell}_{\text{B}}$ line
and, thus, should not be expected to follow the above-mentioned
trend. Note that while in the B phase no quark state is populated,
in both the $C_0$ and $A_0$ phases only the lowest Landau levels
(LLL's) of the $d$- and $u$-quark are. It is also interesting to
notice that for the parametrization $M_0=200$ MeV we are
discussing each first order line ${\ell}_{\text{C}i}$ appears to
be naturally continued by the ${\ell}_{\text{A}i}$ one. As we will
see below this correspondence is not so clear for larger values of
$M_0$. To complete the description of the phase diagram for
$M_0=200$ MeV we present some comments on the second order lines
going from the points $s_i$ to $t_i$, and which separate the
$\mbox{A}_i$ phase from the $\mbox{C}_i$ one. The points on these
lines obey relations that can be obtained by demanding that the
quadratic coefficient of the Landau expansion of Eq.(\ref{omega})
vanishes. Although in general these relations can be only
numerically given, the case for the line connecting $s_0$ to $t_0$
admits a simple analytical expression. It reads
\begin{equation}
\mu = \frac{1}{2^{2/3}}
\sqrt{ \frac{e B}{3 \pi}} \exp{ \left[ 2 \left( 1 - \frac{g_c}{g} \right) \frac{\Lambda^2}{e B} \right] }
\label{muc}
\end{equation}
with $g_c = \pi^2/N_c N_f$. Of course, the equation is valid for
$e B(s_0) < e B < e B(t_0)$. The position of the critical point
$s_0$ can be easily obtained by demanding that Eq.(\ref{muc}) and
the relation $\mu = \frac{2}{3} \ eB$ are simultaneously
satisfied. Note that the latter relation corresponds to the line
${\ell}_{\text{C}1}$ (see Eq.(\ref{va})). On the other hand, to
obtain the location of the critical point $t_0$, Eq.(\ref{muc})
has to be solved together with the relation satisfied by the first
order line separating the $B$ and $C_0$ phases, which follows from
the condition $\Omega\left(\mu,e B, M(0,e B)\right) = \Omega(\mu,e
B, 0)$. This procedure can be generalized so as to determine the
precise position of the rest of the $s_i$ and $t_i$ critical
points. However, due to the lack of simple analytical expressions
for the equations involved, this has to be numerically done.

\begin{figure}[t]
  \includegraphics[width=0.9\linewidth,angle=0]{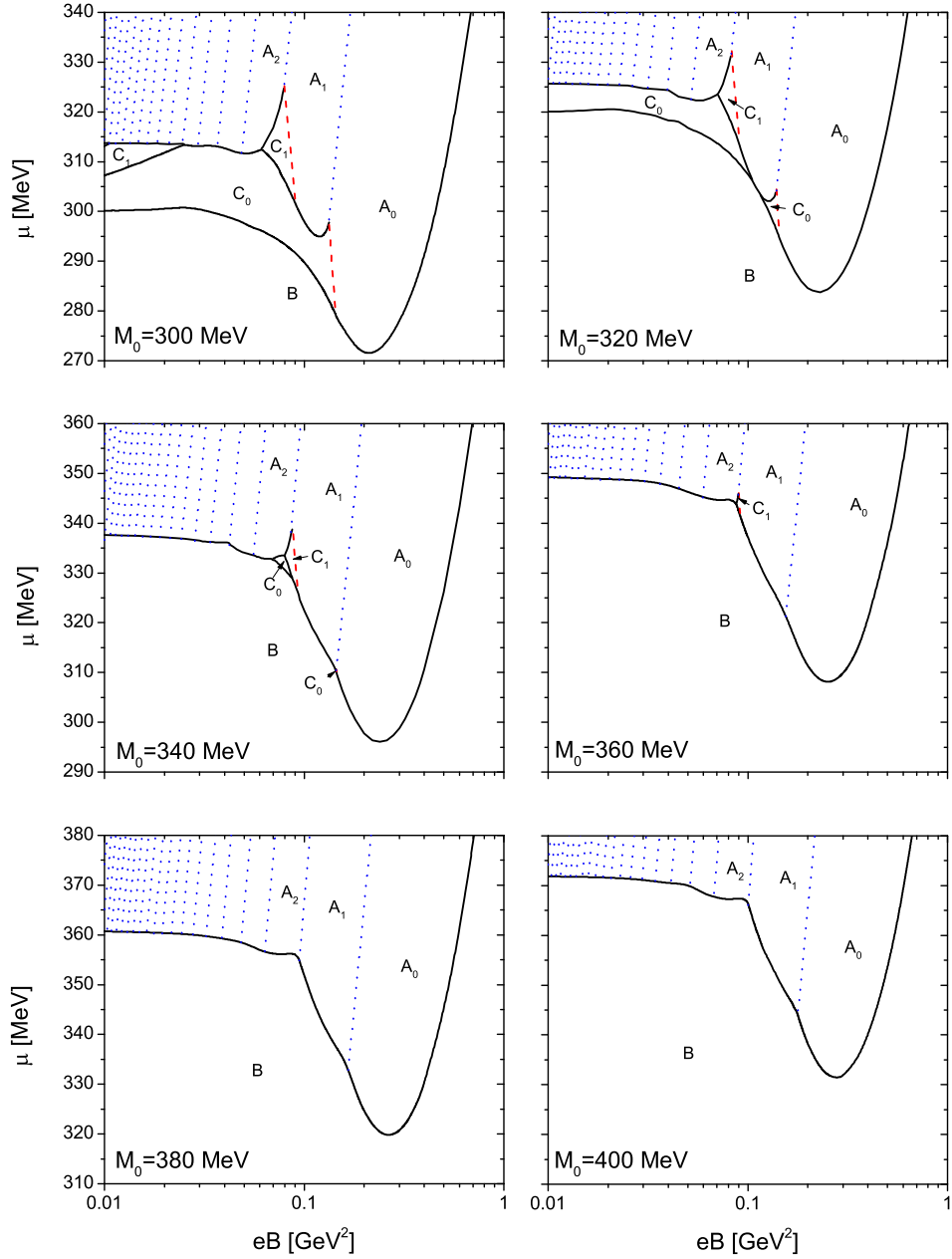}
    \caption{(Color online) Phase diagrams in the $eB-\mu$ plane in the chiral case and
    for various representative values of $M_0$. Full (black) lines represent first order phase transitions
    while dashed (red) lines to second order ones. Dotted (blue) lines correspond to the second
    order transitions which separate the different massless phases.
    Different phases are denoted as in Fig.\ref{fig6}.}
\label{fig8}
\end{figure}

In the rest of this section we discuss how the $e B- \mu$ diagrams
for cold quark matter evolve as we turn to parametrizations
corresponding to the relevant range $300 < M_0 < 600$ MeV.
Diagrams with several values of $M_0$ at intervals of 20 MeV are
shown in Fig.\ref{fig8}. Only parametrizations up to $M_0=400$ MeV
are explicitly displayed. As discussed below, beyond that value of
$M_0$ the corresponding phase diagrams do not involve any
qualitative new feature. Let us consider first the case $M_0=300$
MeV. As we see, there is a considerable simplification with
respect to that of $M_0=200$ MeV. In fact, apart from the ever
present B phase, only two massive phases exist in the relevant
range of magnetic fields (very close to $e B = 0.01$
$\mbox{GeV}^2$ there is a very tiny region of $\mbox{C}_2$ phase
which can hardly be seen in the figure). Thus, contrary to the
case of $M_0=200$ MeV, there is a range of magnetic fields for
which a first order transition can connect the $\mbox{C}_i$ phase
to some phases $\mbox{A}_{i+m}$, with $m > 1$. The fact that the
$\mbox{C}_1$ phase is no longer simply connected can be understood
as due to a "strangulation" of that region caused by the rise of
the central part of the line connecting M to $s_0$. In fact, the
strict correspondence between the first order line
${\ell}_{\text{C}i}$ and the second order one ${\ell}_{\text{A}i}$
(with the same $i$) mentioned above is lost here. Going now to the
case $M_0=320$ MeV, only one sector of the $\mbox{C}_1$ phase (the
one surrounded by the phases $\mbox{C}_0$, $\mbox{A}_1$ and
$\mbox{A}_2$) shows up for $e B > 0.01\ \mbox{GeV}^2$. Moreover,
the $\mbox{C}_0$ phase gets smaller and is split into two pieces.
It is interesting to note that slightly above $e B = 0.1 \
\mbox{GeV}^2$ two first order lines seem to touch at one single
point. In fact this is exactly true for a somewhat lower value
$M_0=319.2$ MeV. In any case, the existence of this particular
meeting point might be worrisome since no more than three first
order lines are expected to converge at one point. However, since
only three different phases ($\mbox{B}$, $\mbox{C}_0$ and
$\mbox{A}_1$) coexist at this point this does not bring any
contradiction with general statistical mechanical arguments. If
$M_0$ is increased further to $M_0=340$ MeV only two small
``islands" of $\mbox{C}_0$ remain: one (which can hardly be seen
in the figure) is separated from the $\mbox{A}_0$ phase by a
second order transition, and the other is fully surrounded by
first order lines. In addition the $\mbox{C}_1$ region gets
somewhat smaller. Note that since for this value of $M_0$ there is
only one single first order transition in the limit of vanishing
magnetic field (see Fig.\ref{fig1}), no other region of any
$\mbox{C}_i$ phase is expected to appear even for lower values of
$e B$. For $M_0=360$ MeV the region $\mbox{C}_1$ becomes very
tiny, and for $M_0=380$ MeV it is not present any more. Note that
from there on (see e.g. the diagram for $M_0=400$ MeV) the
diagrams become very simple displaying only one first order phase
transition for any arbitrary value of the magnetic field
considered. Of course, in addition to it, we have the
${\ell}_{\text{A}i}$ lines separating the different $\mbox{A}_i$
phases. The precise value at which the $\mbox{C}_1$ disappears can
be determined by finding when the points $s_1$ and $t_1$ meet. Of
course, this value coincides with that of $M_0(a)^{hm}$ given in
Eq.(\ref{mhm}). Thus, for our choice, $f_\pi^{ch} = 90$ MeV, the
parametrization beyond which the $e B - \mu$ diagram is
particularly simple corresponds to $M_0=371.46$ MeV. We can also
mention that the $\mbox{C}_0$ phase does not exist for
parametrizations $M_0 > M_0(a)^{lm}$.

\begin{figure}[t]
\hspace*{1cm}
    \includegraphics[width=0.6\linewidth,angle=0]{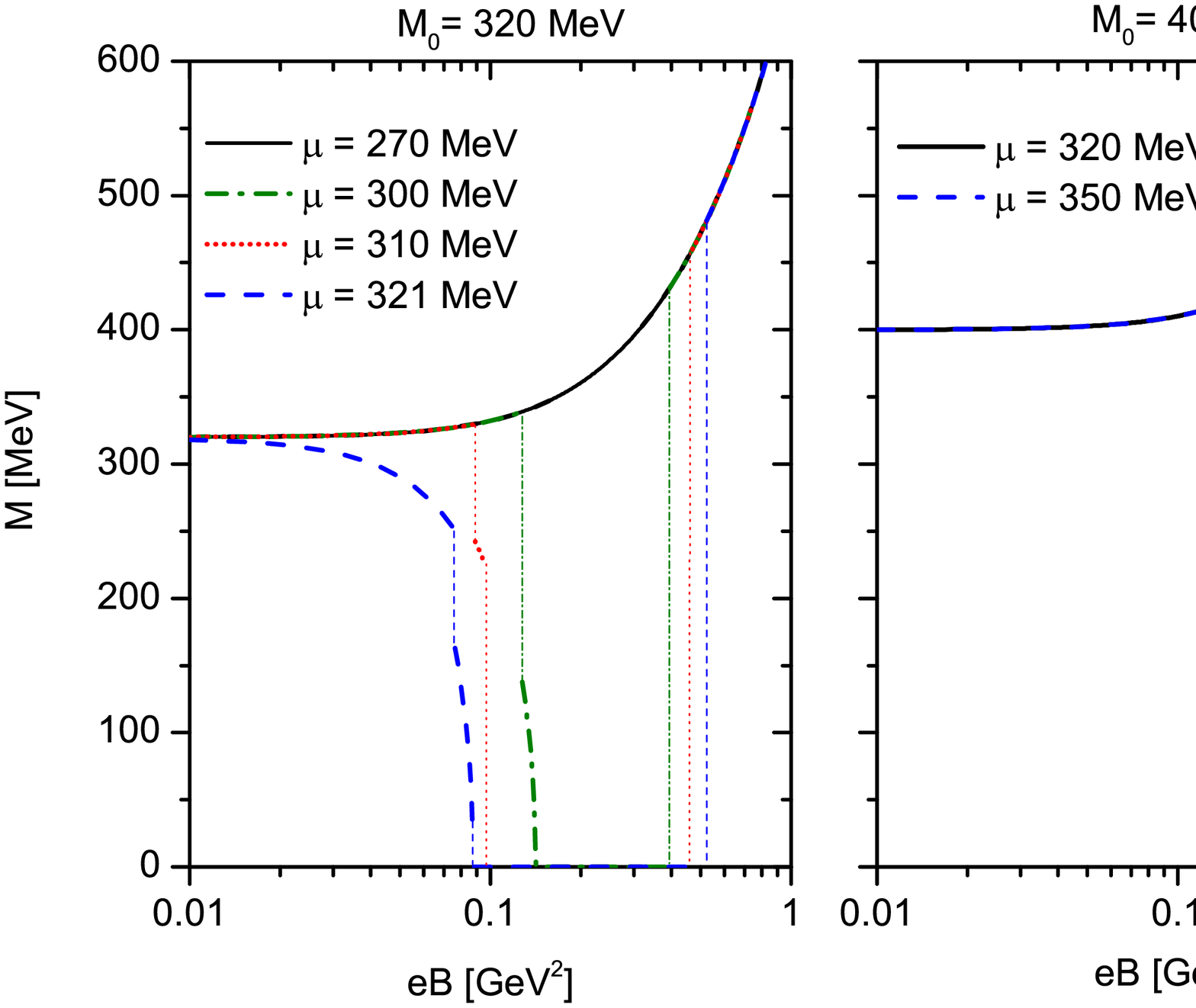}
    \hspace*{\fill}
   \caption{(Color online) Dressed quark mass $M$ as a function of $eB$ in the chiral case
   for several representative values of the chemical potential
   using the parameter sets associated with $M_0= 320$ MeV (left panel) and $M_0 =400$ MeV (right panel).
} \label{fig9}
\end{figure}

It is interesting to address at this point the so-called ``inverse
catalysis effect" recently discussed in the
literature\cite{Preis:2010cq}. This is usually related to a
decrease of the critical chemical potential at intermediate values
of the magnetic fields. Such a phenomenon is clearly observed for
all the cases indicated in Fig.\ref{fig8}. In fact, we see that
after staying fairly constant up to $eB \simeq 0.05\ \mbox{GeV}^2$
the transition line ${\ell}_{B}$ bends down reaching a minimum at
$eB \simeq 0.2 - 0.3\ \mbox{GeV}^2$ after which it rises
indefinitely with the magnetic field. This implies that, in
general, there is some interval of values of the chemical
potential for which an increase of the magnetic field at constant
$\mu$ causes first a transition from the massive phase B to some
massless phase $\mbox{A}_i$ and afterwards from the massless phase
$\mbox{A}_0$ back to massive phase B. This is clearly observed in
Fig.\ref{fig9} where we plot the behavior of $M$ as a function of
$eB$ for several representative values of $\mu$ and $M_0=320$ and
$400$ MeV. An important feature not so often discussed in the
literature (see however Ref.\cite{Boomsma:2009yk} for a brief
comment on this) can also be noticed in the case of $M_0=320$ MeV:
when the system is in a $C_i$ phase there is an actual ``inverse
catalysis effect" in the sense that the order parameter for
spontaneous chiral symmetry breaking ($M$ or the chiral
condensate) does decrease with the magnetic field while staying in
the same phase. For example, in the case of $\mu=300$ MeV (green
dot-dashed line) there is first a "catalysis effect" while the
system stays in the B phase, then at about $eB=0.127\
\mbox{GeV}^2$ there is a first order transition to the
$\mbox{C}_0$ phase after which the "inverse catalysis effect" can
be clearly observed. This situation proceeds up to $eB=0.139\
\mbox{GeV}^2$ where there is a second order transition to the
$\mbox{A}_0$ phase. Eventually, at $eB=0.394\  \mbox{GeV}^2$ the
system undergoes a new first order transition that brings it back
to the B phase. In the case of $\mu=310$~MeV (red dotted line) the
situation is similar except for the fact that the intermediate
transition is of first order. Finally, for $\mu=321$~MeV (blue
dashed line) the system is in the $\mbox{C}_0$ phase even for very
small magnetic fields and, thus, the ``inverse catalysis effect"
is already present at low values of $eB$. At $eB=0.074
\mbox{GeV}^2$ there is a first order transition to the
$\mbox{C}_1$ phase after which the ``inverse catalysis effect" can
still be clearly observed. At $eB=0.088\ \mbox{GeV}^2$ there is a
second first order transition to the $\mbox{A}_1$ where $M$
vanishes and, finally, at $eB=0.39\ \mbox{GeV}^2$ there is a new
first order transition to the B phase. Note that between the last
two transitions there is a second order transition from the
$\mbox{A}_1$ phase to the $\mbox{A}_0$ one which, of course, in
the present chiral case does not produce any effect on the
behavior of $M$ as a function of $eB$.

We conclude this section with a brief comment on the sometimes used LLL approximation.
It is clear that such an approximation is well justified if only such Landau level is
involved in the transitions under study. For example, for the parametrization
$M_0=300$ MeV, this is the case for the lowest $\mu$ first order transition
and the whole range of values of $eB$ considered. However, the situation changes
as $M_0$ increases. Already for $M_0 = 360$ MeV it can only be safely used
to describe the transition between the $B$ and $C_0$ phases, i.e. for rather
large values of $e B$.

\section{Finite current quark masses }

The addition of a non zero current mass to the problem brings
along a few qualitative and quantitative differences. To begin
with, there is no longer a universal character to the phase
diagram: parameter sets associated with different values of
$f_\pi$ are not related among themselves through a scale change.
In the rest of this work, we set $m_\pi=138$ MeV and $f_\pi=92.4$
MeV and choose a value of $M_0$ within the phenomenological range
$300 \lesssim M_0 \lesssim 600$ MeV in order to fix the model
parameters  $m_c$, $g=G\Lambda^2$ and $\Lambda$. The resulting
values as well as those associated with the corresponding chiral
condensates are given in Appendix B.

For zero magnetic field, the $M_0-\mu$ phase diagram is
qualitatively similar to the one corresponding to the chiral case
(see Fig.1) except for the fact that the highest $\mu$ second
order transitions occurring for $M_0 < M_0(b)$ become smooth
crossovers here. For the values of $f_\pi$ and $m_\pi$ given
above, we find $M_0(a)=361.2$ MeV and $M_0(b) = 300.1$ MeV.

\begin{figure}[t]
\hspace*{1.5cm}
    \includegraphics[width=0.6\linewidth,angle=0]{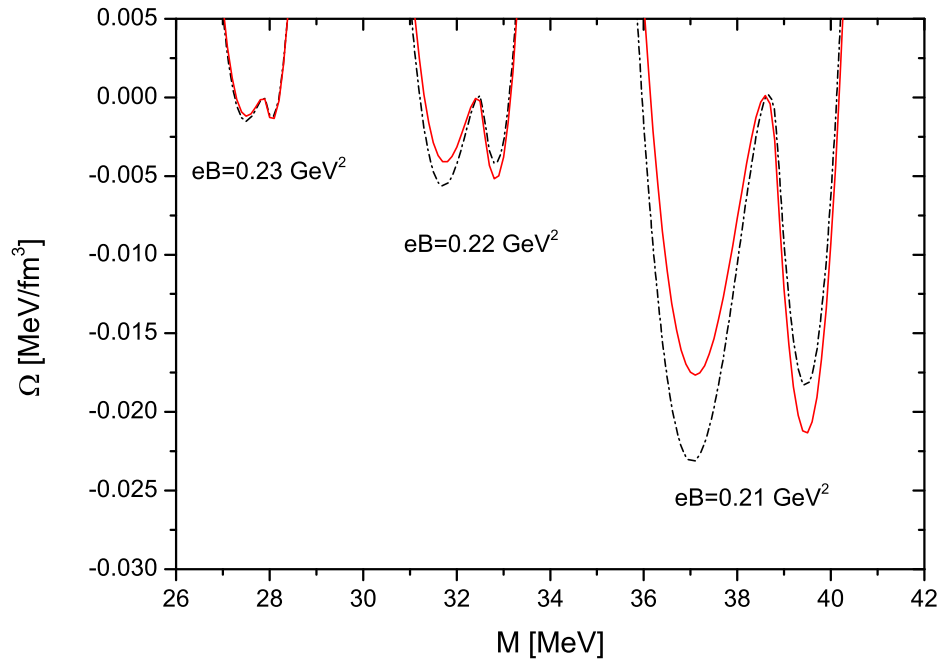}
    \hspace*{\fill}
    \caption{(Color online) Thermodynamical potential as a
function of $M$ for some representative values of $eB$. In each
case we have subtracted the value of the thermodynamical potential
at the intermediate maximum so as to be able to include all the
cases in the same plot. Red full lines (black dot-dashed lines)
correspond to a value of chemical potential slightly above (below)
the critical value.} \label{fig10}
\end{figure}

We turn now to the case of finite magnetic field. Before
presenting the actual phase diagrams we will discuss the main
qualitative differences introduced by the existence of finite
current quark mass. We start by the second order lines which
separate the different $\mbox{A}_i$ phases in the chiral case, and
whose equations are given in Eq.(\ref{va}). Let us recall that the
corresponding critical chemical potentials are the values at which
new LL's contribute to the sum in the last line of
Eq.(\ref{omega}) for $M=0$. In the case of finite quark masses,
although $M$ never vanishes we can still define the value of the
chemical potential at which new LL's are populated, i.e. the one
that satisfies the condition $\mu = \sqrt{M^2 + 2\ k\ c_f\ eB/3
}$. As it turns out, in all the cases under study we found that
for a given value of $eB$ there is no second order transition
located at this chemical potential but a (weak) first order one in
its vicinity, the transition becoming weaker as the critical $\mu$
increases. Namely, the second order ${\ell}_{\text{A}i}$ lines
present in the chiral case become first order here, being
signalled by very small jumps in the dressed mass. The situation
is illustrated in Fig.~\ref{fig10} where we plot the
thermodynamical potential as a function of $M$ for some
representative values of $eB$. In each case, we have chosen one
value of $\mu$ above the critical chemical potential and the other
one below. Moreover, in each case we have subtracted the value of
the thermodynamical potential at the intermediate maximum so as to
be able to include all the cases in the same plot.
Fig.~\ref{fig10} clearly displays the existence of two solutions
on either side of the point at which the condition mentioned above
is satisfied, and how one of them becomes the absolute minimum
depending on whether $\mu$ is below (black full line) or above
(red dot-dashed line) its critical value. The decrease of the jump
in mass as $eB$ (and, thus, the critical chemical potential)
increases can also be observed in Fig.~\ref{fig10}. Since for
finite current quark masses the different $\mbox{A}_i$ phases are
separated by first order lines (in the same way as different
$\mbox{C}_i$ phases are), it is no longer possible to distinguish
between ${\ell}_{\text{A}i}$ and ${\ell}_{\text{C}i}$ lines as
done in the chiral case: one simply has a single continuous first
order line that plays their role.

\begin{figure}[t]
\hspace*{1cm}
\includegraphics[width=0.7\linewidth,angle=0]{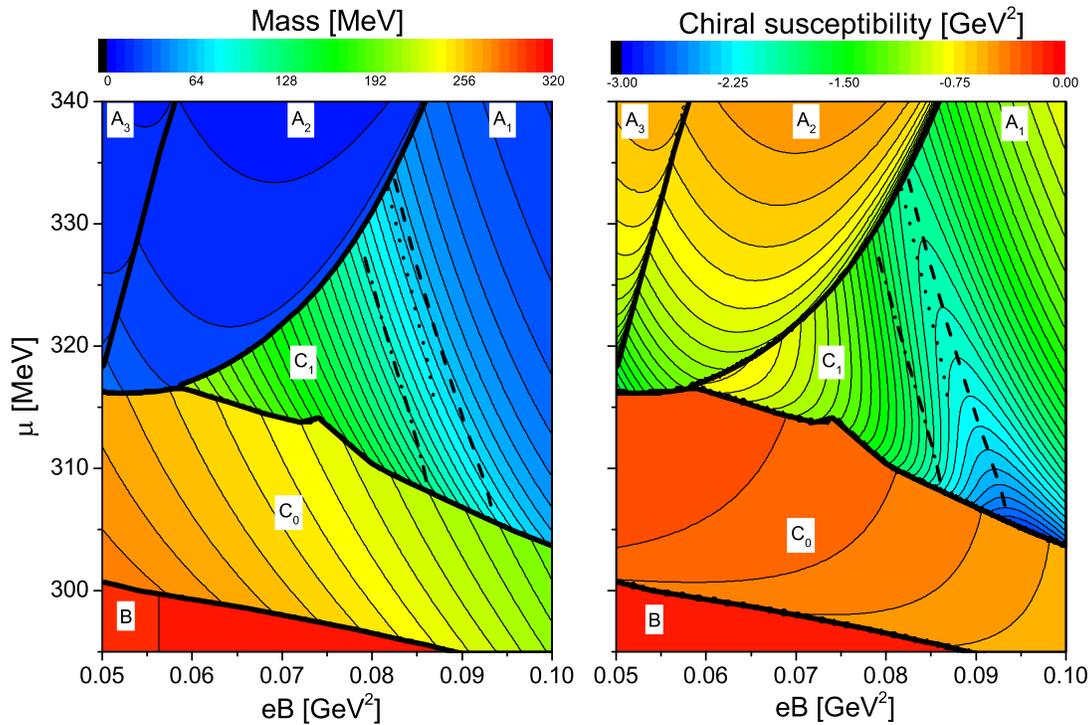}
    \hspace*{\fill}
    \caption{(Color online) Contour plots of the dressed mass (left panel) and
    the chiral susceptibility (right panel) as functions of $\mu$ and
    $eB$. Parameters are as in the chiral case for $M_0=300$ MeV but
    with $m_c = 1$ MeV. Full (black) lines represent first order phase transitions.
    The different definitions discussed in the text are used to obtain the
    crossover transition lines: Def. $i)$ (peak of $dM/dB$) is represented by
    dot-dashed line, Def. $ii)$ (peak of $\chi_{ch}$ as a function of $eB$) by
    a dashed line and Def. $iii)$ (peak of $\chi_{ch}$ as a function of $\mu$) by
    a dotted line. Def. $iv)$ corresponds to the line of ``slowest descent" from the
absolute peak of the chiral susceptibility (dark blue region in
right panel) which basically coincides with that of Def. $ii)$.
Different phases are denoted as in Fig.\ref{fig6}.} \label{fig11}
\end{figure}

We turn now to the fate of the transition lines that separate the
phases $\mbox{C}_i$ from the $\mbox{A}_i$ ones, and which are of
second order in the chiral case. For finite current quark masses
these transitions become smooth crossovers. Consequently, as
already mentioned in Sec.III, there is not a unique way to define
their position. In the present case, considering the peak of the
derivative of $M$ with respect to $\mu$ or $B$ gives rise to two
possible prescriptions. As it happens, however, due to the
particular form of the transition lines (rather parallel to the
$\mu$ axis as one can expect from the chiral case, see
Fig.~\ref{fig8}) we find that in general there is no peak of
$dM/d\mu$. Thus, we are only left with the second possibility that
we denote Def. $i)$. For basically the same reason, the transition
line defined as the position of the peaks of the chiral
susceptibility when plotted as a function of $eB$ at fixed $\mu$
(denoted Def. $ii)$) do not coincide, in general, with the one
that follows from the alternative possibility (denoted Def.
$iii)$), i.e. the peaks of the chiral susceptibility when plotted
as a function of $\mu$ at fixed $eB$. Moreover, in the latter case
the transition line tends to be washed away when the current quark
mass is varied from $m_c=0$ to its corresponding physical value.
To avoid this dependence on the somewhat ad-hoc chosen direction
in the $eB-\mu$ plane one can define the transition line as the
ridge occurring in the chiral susceptibility when regarded as a
two dimensional function of $eB$ and $\mu$. Mathematically, it can
be defined by using for each value of the susceptibility (starting
from its maximum value in the given region) the location of the
points at which the gradient in the $eB-\mu$ plane is smaller. We
denote this as Def. $iv)$. The situation is illustrated in
Fig.\ref{fig11} where we plot the contour lines corresponding to
$M$ (left panel) and the chiral susceptibility (right panel) for
the case in which $\Lambda$ and $g$ take the values associated to
the chiral case with $M_0=300$ MeV but $m_c$ is arbitrarily set to
$m_c = 1$ MeV. While non physical, for this parameter set it is
possible to use all the alternative ways to define the transition
lines mentioned above (in particular, that associated with the
peak of $d\chi_{ch}/d\mu$ which rapidly disappears as $m_c$
increases). In this way one can obtain some measure of the
ambiguity this introduces in the location of the transition line.
In Fig.\ref{fig11} the thick dot-dashed lines correspond to Def.
$i)$, the dashed line to Def. $ii)$ and the dotted line to Def.
$iii)$. From the left panel of this figure it is quite clear that
the latter definition leads to a transition line which basically
coincides with the one obtained from Def $iv)$, which in turn
corresponds to the line of ``slowest descent" from the absolute
peak of the chiral susceptibility (darker blue region). Although
some alternative definitions are still possible (e.g. the contour
line in the susceptibility diagram that, at the contact point, is
tangent to the first order line that separates the $\mbox{C}_1$ or
$\mbox{A}_1$ phases from the $\mbox{A}_2$) we see that the
different definitions lead to qualitatively similar results. Thus,
in what follows we will use Def. $iv)$ having in mind that to
ensure the real existence of the transition line one should be
able to define it in at least more than one way. This requires
that Def. $iv)$ must be complemented with the condition that on
each side of the curve there should exist at least one region such
that there is a maximum in the susceptibility for an arbitrary
path connecting both regions. As a corollary of this discussion we
note that the location of the critical points equivalent to the
points $s_i$ and $t_i$ discussed in the previous section is also
subject to definition ambiguities.

\begin{figure}[t]
    \includegraphics[width=0.8\linewidth,angle=0]{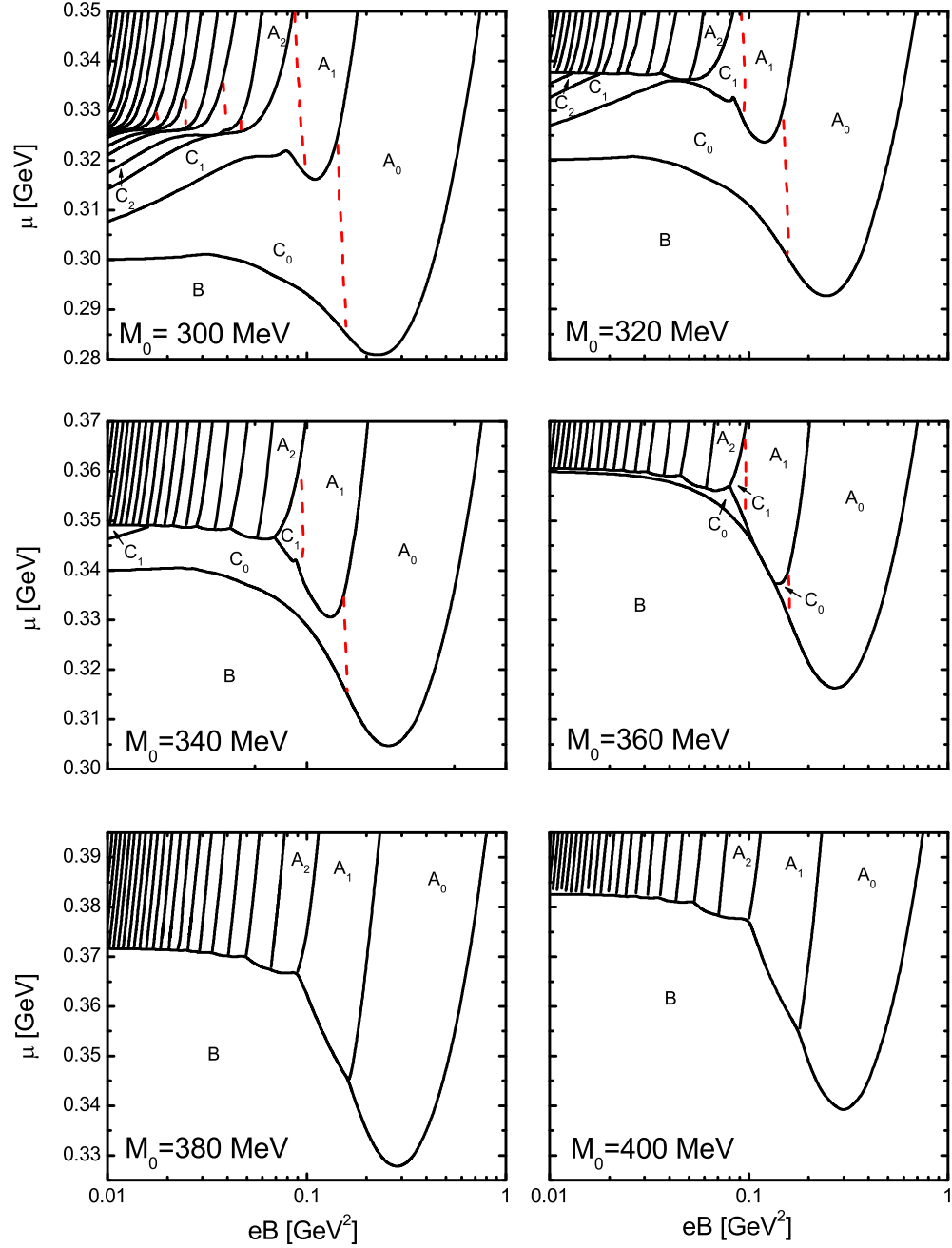}
    \caption{(Color online) Phase diagrams in the $eB-\mu$ plane in the case of finite current quark masses and
    for various representative values of $M_0$. Full (black) lines represent first order phase transitions
    while dashed (red) lines crossover ones. Different phases are denoted as in Fig.\ref{fig6}.}
\label{fig12}
\end{figure}

The phase diagrams in the $eB-\mu$ plane for different values of
$M_0$ are presented in Fig.\ref{fig12}. Apart from the particular
features just discussed, we observe that the general trend is
similar to the one of the chiral case shown in Fig.\ref{fig8}: For
low $M_0$ values there are several different transitions that
coalesce into fewer transitions as $M_0$ is increased. Moreover,
while several crossovers between $\mbox{C}_i$ and $\mbox{A}_i$
phases are still present for the set $M_0=300$ MeV, they continue
to exist from $M_0=320$ MeV until $M_0=360$ MeV only for $i=0,1$.
It would seem that a phase diagram for a given $M_0$ in the chiral
case is always similar to another one in the non chiral case with
larger $M_0$. In particular, for finite quark masses the value of
$M_0$ above which there is a unique transition is $375.9$ MeV. It
interesting to note that in this case the $\mbox{C}_0$ phase (in
particular the piece completely surrounded by first order
transition lines) is the last one to disappear. In fact, the
$\mbox{C}_1$ phase ceases to exist for values of $M_0$ slightly
above $360$ MeV.

\begin{figure}
\includegraphics[width=0.4\linewidth]{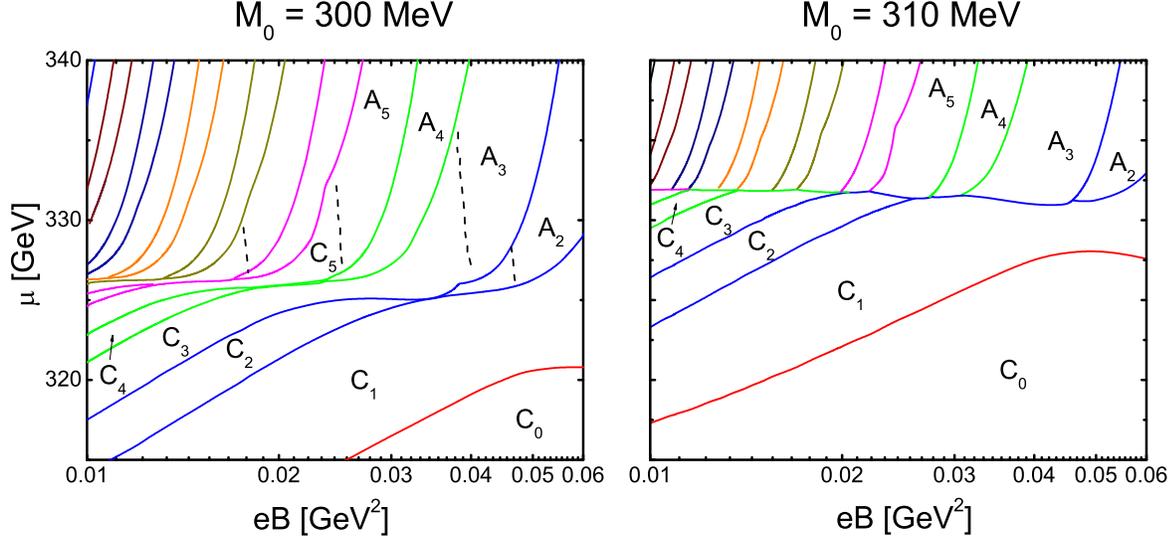}
\hspace*{\fill} \caption{(Color online) Detail of the $eB-\mu$
phase diagram in the case of finite current quark masses for $M_0=
300$ MeV (left panel) and $M_0 = 310$ MeV (right panel). Different
phases are denoted as in Fig.\ref{fig6}.}
 \label{fig13}
\end{figure}

The way in which curves merge together as $M_0$ increases is
qualitatively similar to the chiral case. This is shown in
Fig.\ref{fig13} where we display a detail of the $eB-\mu$ diagram
for $M_0 = 300$ MeV (left panel) and $M_0 = 310$ MeV (right
panel). We see that the curves develop a flat cubic-like region,
through which they come into contact. A transition curve that is
nearly independent of the chemical potential is eventually formed
from these flat regions. From the original curves, the vertical
parts with higher chemical potential (i.e. those separating the
$\mbox{A}_i$ phases) continue to exist as $M_0$ is increased,
while the lower chemical potential parts of the curves tend to
move to lower magnetic field values and eventually disappear. In
Fig.\ref{fig13} it is also quite clearly seen that the curves join
in pairs, transition curves being colored in the figure as to
indicate which curves join between themselves. For example, the
first line to the right, corresponding to the simultaneous
population of the second $d$-quark LL and the first $u$-quark LL,
merges with the following curve which corresponds to the
population of the third $d$-quark LL.

\begin{figure}[t]
\hspace*{1cm}
\includegraphics[width=0.6\linewidth,angle=0]{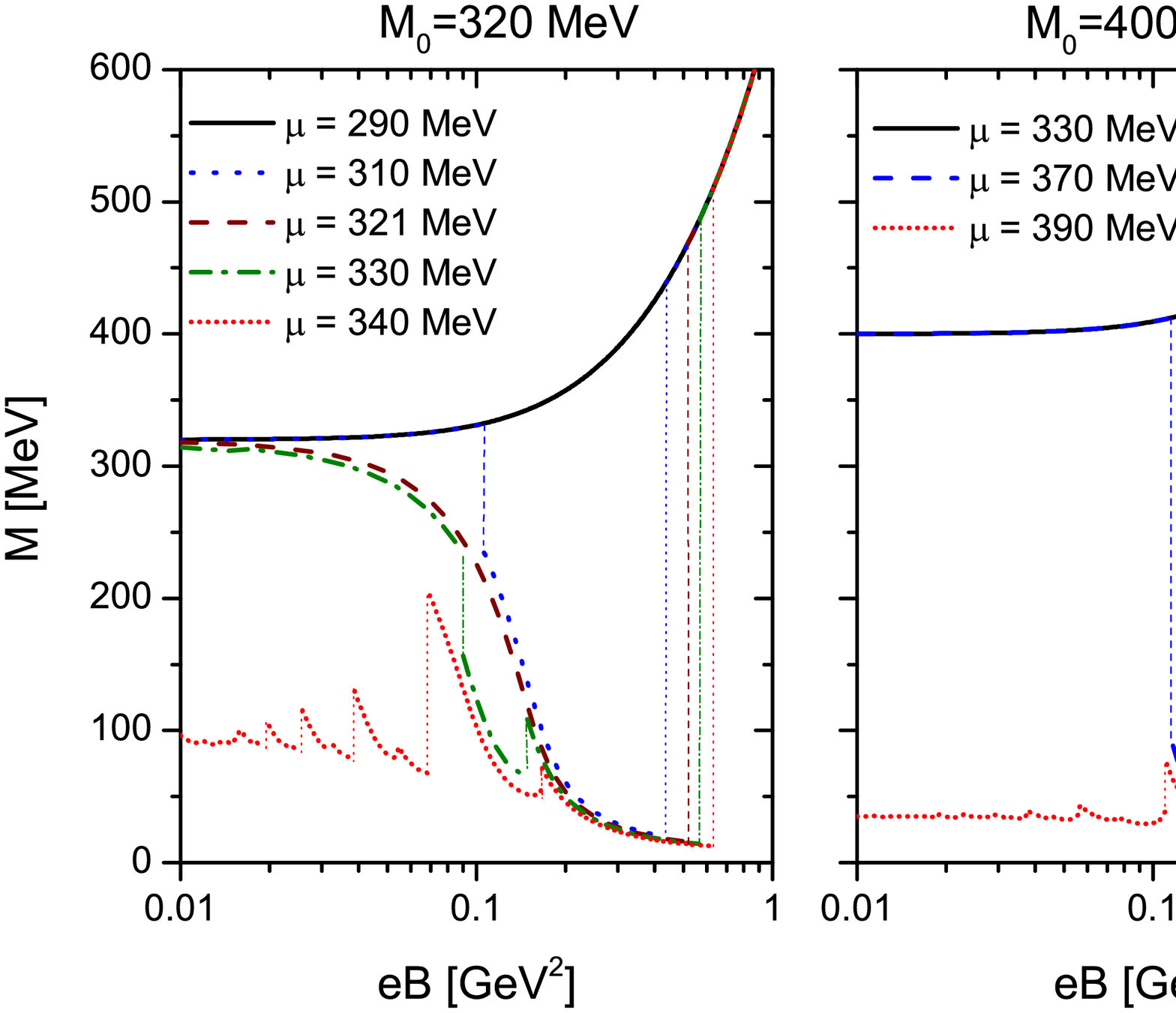}
\hspace*{\fill} \caption{(Color online) Dressed quark mass $M$ as
a function of $eB$ for several representative values of the
chemical potential using the parameter sets associated with $M_0=
320$ MeV (left panel) and $M_0 =400$ MeV (right panel). }
\label{fig14}
\end{figure}

We end this section by discussing the ``inverse catalysis effect"
for the case in which a finite current mass is present. In
Fig.\ref{fig14}, we display the behavior of the mass as a function
of magnetic field for several chemical potentials, and the
$M_0=320$ MeV and $M_0=400$ MeV parameter sets. The complex phase
structure for the $M_0=320$ MeV case accounts for the different
possible behaviors depending on the chemical potential. For
$\mu=290$ MeV, the system is in the B phase for  the whole range
of magnetic fields, and the catalysis effect is clearly seen. For
$\mu=310$ MeV, a similar behavior is seen, except for a middle
section where the system passes through a $\mbox{C}_0$ phase and
an $\mbox{A}_0$ phase before returning to the vacuum phase again.
As opposed to the chiral case, the transition from $\mbox{C}_0$ to
$\mbox{A}_0$ is not particularly noticeable since at most the
transition is signaled by a peak in the susceptibility or the
derivatives in the order parameter as already discussed. In this
region of the curve, as well as in the rest of the following
curves, the effect of inverse catalysis is also present.  In fact,
it is absolutely dominant except for barely noticeable regions in
the $\mbox{A}_i$ phases for the $\mu=340$ MeV curve, and we can
conclude that within phases with non-zero quark density, the value
of $M$ is typically a decreasing function of the magnetic field,
while catalysis occurs principally in the vacuum phase. In
particular, for $\mu=321$ MeV, the phase remains in $\mbox{C}_0$
for a significant range of magnetic fields and the mass decreases
continuously. Paying attention to the cases $\mu = 330$ MeV and
$\mu = 340$ MeV, we will also note that when we move to a phase of
increasing $i$, so as to populate new LL's, the discontinuity will
be towards a lower mass, while if $i$ decreases in the transition
so as to leave a formerly occupied Landau level empty, the jump
will be towards a higher mass.

\section{Summary and conclusions}

 In this work we have considered the phase structure of magnetized cold
 quark matter in the framework of the two-flavor Nambu-Jona-Lasinio models.
  As is well known, even in the simplest version of these models there is a
  rather broad range of phenomenologically acceptable values for the
  corresponding model parameters. Thus, we have performed a detailed analysis
  of how the character and location of the different types of phase transitions
  depend on the chosen parametrization. As frequently done in the literature,
  we have specified each parametrization by the associated value of the dressed
  quark mass in the vacuum at vanishing magnetic field $M_0$, with the
  phenomenological range given by $300 \lesssim M_0 \lesssim 600$ MeV \cite{Buballa:2003qv}.
  We have first discussed the simpler situation in which the chiral limit is taken.
  In this case the phase structure is basically dictated by the ratio $M_0/f_\pi^{ch}$.
  For $M_0/f_\pi^{ch} > 4.127$ such structure is particularly simple since only one single
  first order transition line ${\ell}_{\text{B}}$ exists. This line separates the vacuum phase B
   from the ones in which a certain number of Landau levels associated with massless $u$- and $d$-quarks
   are populated. Following Ref.\cite{Ebert:1999ht} we denote the latter ones as $\mbox{A}_i$ phases.
   They are separated by second order transition lines that we called ${\ell}_{\text{A}_i}$.
   On the other hand for $M_0/f_\pi^{ch} < 4.127$ the phase diagram is more complex since
   additional first phase order and second order transition lines appear as $M_0$ decreases.
   In particular, there appear new phases $\text{C}_i$ in which chiral symmetry
   is only partially restored. Namely, for a given magnetic field the corresponding dressed mass
   is smaller than its vacuum value and depends on the chemical potential. It is important to stress that,
   for a typical value $f_\pi^{ch} = 90$ MeV, the parametrization below which these new phases and transition
   lines appear corresponds to $M_0 = 371.46$ MeV, a value which is well inside the phenomenological
   acceptable range quoted above.

 When a finite current mass is included in the model there are some changes but the general structure
 of the $eB-\mu$ phase diagram remains the same, with a particular $M_0$ diagram in the chiral case
 typically very similar to another one with larger $M_0$ in the non chiral case. In particular, a
 slightly higher value $M_0=375.9$ MeV is required for the passage from the broken symmetry phase
 to the restored phases $\mbox{A}_i$
 to occur in one single transition.  As in the chiral case, below this critical $M_0$ value several
 transitions are needed to move from the vacuum phase B to the chiral symmetry restored phases $\mbox{A}_i$, if chemical
  potential is increased at constant magnetic field.
One of the most notable qualitative modifications induced by the
presence of a finite
 quark mass is related to the character of the transitions between the $\mbox{A}_i$ phases:
 while in the non chiral case the transitions
 are of
 first order, signaled by a jump both in the density and the dressed mass, in the chiral case
 the order parameter is zero in all of these phases and the transition is second order and
 signaled by a discontinuous derivative of the quark density. The transitions between a
 $\mbox{C}_i$
  phase and the corresponding  $\mbox{A}_i$ phase are also different, being second order transitions in the
  chiral case and smooth crossovers when current mass is finite. Several definitions for the location
  of the crossover transitions
  were studied, finding in general that even though the
different definitions introduce certain
  ambiguity as to the exact location of the transition, in all studied cases they agree on whether
  the transition actually exists or not. As well, in what respects to their tendency to disappear
  as $M_0$ increases, these crossovers behave similarly to their second order analogues occurring in the chiral case.

The behavior of the dressed mass for a constant $\mu$ in response
to magnetic field was studied as well for both the chiral and non
chiral cases, resulting in different effects depending on the
phase. On the one hand, the increase in dressed mass with magnetic
field, known as magnetic catalysis, was principally seen in the
vacuum phase B, where symmetry is fully broken, in both chiral and
non chiral cases. On the other hand, phases with non zero quark
density and finite dressed mass ($\mbox{C}_i$ phases in the chiral
case, and $\mbox{C}_i$ and $\mbox{A}_i$ in the non chiral case)
showed a dominant decrease in the dressed mass as magnetic field
increased. This can be taken as a manifestation of ``inverse
magnetic catalysis" usually associated with a decrease of the
critical chemical potential at intermediate values of the magnetic
fields\cite{Preis:2010cq}. It should be noted that these
continuous drops in the mass occurred within single phases, and
that discontinuous jumps occurred whenever a new phase with a
different amount of occupied Landau levels was reached.

Throughout this work only the simplest version of the two flavor
NJL with maximum flavor mixing has been considered. It is clear
that the parametrization dependence of the phase structure of
magnetized cold quark matter as described by possible extensions
of the model which incorporate the effect of different amounts of
flavor mixing\cite{Boomsma:2009yk}, color superconductor
channels\cite{Fayazbakhsh:2010gc,Ferrer:2012wa}, vector
interactions\cite{Denke:2013gha}, strangeness degrees of
freedom\cite{Menezes:2009uc}, etc is interesting and certainly
deserves further investigation. Of course, the analysis of how the
phase structure of magnetized quark matter at finite temperature
depends on the model parametrization should also be
addressed\cite{Allen:2013eha}. In this respect, however, it is
important to mention that the model extensions that incorporate
the effect of the Polyakov loop reduce, in the low temperature
region, to the type of model studied here.

\section*{ACKNOWLEDMENTS}

Fruitful discussions with Marcus B. Pinto, Debora P. Menezes and
Daniel Gomez Dumm are greatly acknowledged. This work has been
partially funded by CONICET (Argentina) under grants PIP 00682 and
by ANPCyT (Argentina) under grant PICT-2011-0113.

\section*{APPENDIX A: Parametrization in the chiral case}

\newcounter{erasmo}
\renewcommand{\thesection}{\Alph{erasmo}}
\renewcommand{\theequation}{\Alph{erasmo}.\arabic{equation}}
\setcounter{erasmo}{1} \setcounter{equation}{0} 

In this Appendix we provide some details of the way in which the
parameters are determined in the chiral limit. In this case the
model has only two parameters: the coupling constant $G$ and the
3D cutoff $\Lambda$. In order to work along ``a line of constant
physics" we choose to determine them so as to reproduce a certain
value of $f_\pi^{ch}$, taking the dressed mass $M_0$ as a free
parameter which takes values within a typical range $200-600$ MeV.
The set of equations to be satisfied by the dimensionless coupling
$g= G \Lambda^2$ and the cutoff $\Lambda$ are the $T=\mu=0$ gap
equation
\begin{equation}
g_c = g\ f\left(\frac{M_0}{\Lambda}\right)
\end{equation}
together with
\begin{equation}
\left(f_\pi^{ch}\right)^2 = \frac{N_c}{2 \pi^2}  \ \Lambda^2
\left[ \frac{M_0^2}{\sqrt{M_0^2 + \Lambda^2}} -
f\left(\frac{M_0}{\Lambda}\right) \right] \label{fpich}
\end{equation}
The second equation corresponds to the expression for $f_\pi$ in
the chiral limit. Moreover, $g_c = \pi^2/(N_c N_f)$ is the
critical dimensionless coupling above which the gap equation has
non-trivial solutions and
\begin{equation}
f(x) = \sqrt{1+x^2} - x^2 \ln{ \left( \frac{1 + \sqrt{1 +
x^2}}{x}\right) }
\end{equation}

\begin{figure}[t]
\hspace*{2cm}
\includegraphics[width=0.8\linewidth,angle=0]{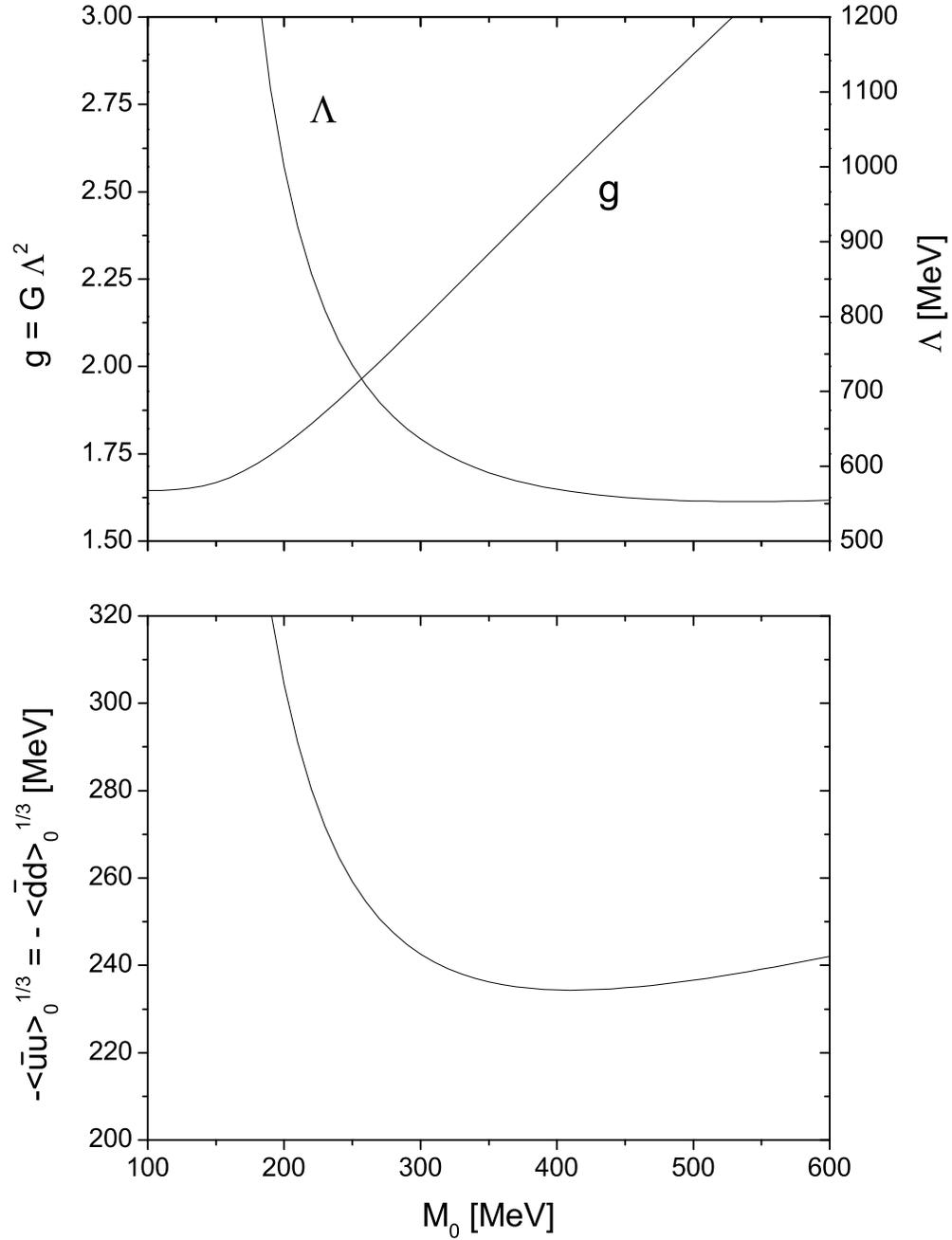}
\hspace*{\fill} \caption{Upper panel: Cutoff parameter $\Lambda$
and dimensionless coupling constant $g$ as functions of the
dressed quark mass $M_0$. Lower panel: Chiral quark condensate as
a function of the dressed quark mass $M_0$.  } \label{figapp}
\end{figure}

The numerical results for $g$ and $\Lambda$ as functions of $M_0$
are shown in the upper panel of Fig.\ref{figapp}. In the lower
panel we display the values of the quark condensates associated
with the corresponding values of the parameters. Here, we have
chosen a typical value for $f_\pi^{ch} = 90$ MeV. Note, however,
that since in the chiral limit $f_\pi^{ch}$ is the only
dimensionful quantity in the problem any dimensionful quantity
(expressed in natural units) has to be the product of some
$f_\pi^{ch}$-independent constant multiplied by some power of it.
This means that if,
 for example, in the lower panel of
Fig.\ref{figapp} we divide the quantities in both axes by
$f_\pi^{ch} = 90$ MeV the resulting curve is universal in the
sense that it does not depend on the chosen value of $f_\pi^{ch}$.
Of course, some extra dependence on the procedure used to
regularize the UV divergencies might still exist.

\section*{APPENDIX B: Parametrization in the finite quark case}

In this Appendix we give the model parameters used in our
calculations of Sec. IV, i.e. for the non-chiral case. They are
listed in Table I. As stated in the main text the are determined
so as to reproduce the physical values $m_\pi~=~138.0$ MeV and
$f_\pi = 92.4$ MeV for a chosen value of dressed quark mass $M_0$
within the phenomenological range $300 \lesssim M_0 \lesssim 600$
MeV \cite{Buballa:2003qv}. The resulting values of the condensates
$- <\bar u u>_0^{1/3}$ = $-<\bar d d>_0^{1/3}$ are also given. We
remind here that the limits extracted from sum rules are
$190~\mbox{MeV} <$ $- <u \bar u>_0^{1/3}~<~260$ MeV at a
renormalization scale of 1 GeV\cite{Dosch:1997wb}, while typical
lattice calculations yield $- <u \bar u>_0^{1/3} = 231 \pm 8 \pm
6$ MeV \cite{Giusti:1998wy} (see e.g. Ref.\cite{McNeile:2005pd}
for some other lattice results).

\vspace*{3cm}

\begin{table}[h]
\centering
\begin{tabular}{ccccc}
\hline \hline
 \hspace{0.5cm} $M_0$  \hspace{0.5cm}  & \hspace{0.5cm}  $m_c$  \hspace{0.5cm}  &
  \hspace{0.7cm} $g=G\Lambda^2 $  \hspace{0.7cm}  &  \hspace{0.5cm}  $\Lambda$   \hspace{0.5cm} &
   \hspace{0.5cm}  $-<\bar u u>_0^{1/3}$ \hspace{0.5cm}  \\
\hline
  MeV   &  MeV  &      &    MeV     & MeV                           \\
\hline
\ 300\ & \ 5.175\ & \ $2.062$ \ & \ 664.4\ & \ 250.8 \ \\
\ 310\ & \ 5.307\ & \ $2.099$ \ & \ 651.0\ & \ 248.7 \ \\
\ 320\ & \ 5.419\ & \ $2.136$ \ & \ 639.5\ & \ 246.9\ \\
\ 340\ & \ 5.595\ & \ $2.212$ \ & \ 620.9\ & \ 244.3\ \\
\ 360\ & \ 5.716\ & \ $2.288$ \ & \ 606.8\ & \ 242.5\ \\
\ 380\ & \ 5.792\ & \ $2.364$ \ & \ 596.1\ & \ 241.4\ \\
\ 400\ & \ 5.833\ & \ $2.440$ \ & \ 587.9\ & \ 240.9\ \\
\hline \hline
\end{tabular}
\caption{Parameter sets for the non chiral case.} \label{table1}
\end{table}

\pagebreak

\end{document}